\documentclass[a4paper,fleqn,usenatbib]{mnras}

\usepackage[T1]{fontenc}
\usepackage{ae,aecompl}

\usepackage{graphicx}	
\usepackage{amsmath}	
\usepackage{amssymb}	
\usepackage[flushleft]{threeparttable} 
\usepackage[caption=false]{subfig} 

\newcommand{\Nant}{N_\textrm{A}}
\newcommand{\Ngrid}{N_\textrm{g}} 

\newcommand{\dif}{\mathrm{d}}

\title[E-field Parallel Imaging Correlator]{A Generic and Efficient E-field Parallel Imaging Correlator for Next-Generation Radio Telescopes}

\author[Thyagarajan et al.]{
Nithyanandan Thyagarajan,$^{1}$\thanks{E-mail: t\_nithyanandan@asu.edu}
Adam P. Beardsley,$^{1}$
Judd D. Bowman$^{1}$
\newauthor
and Miguel F. Morales$^{2}$
\\
$^{1}$Arizona State University, School of Earth and Space Exploration, Tempe, AZ 85287, USA\\
$^{2}$University of Washington, Department of Physics, Seattle, WA 98195, USA\\
}

\date{Accepted XXX. Received YYY; in original form ZZZ}

\pubyear{2015}

\begin{document}
\label{firstpage}
\pagerange{\pageref{firstpage}--\pageref{lastpage}}
\maketitle

\begin{abstract}
Modern radio telescopes are favouring densely packed array layouts with large numbers of antennas ($\Nant\gtrsim 1000$). Since the complexity of traditional correlators scales as $\mathcal{O}(\Nant^2)$, there will be a steep cost for realizing the full imaging potential of these powerful instruments. Through our generic and efficient E-field Parallel Imaging Correlator (EPIC), we present the first software demonstration of a generalized direct imaging algorithm, namely, the Modular Optimal Frequency Fourier (MOFF) imager. Not only does it bring down the cost for dense layouts to $\mathcal{O}(\Nant\log_2\Nant)$ but can also image from irregular layouts and heterogeneous arrays of antennas. EPIC is highly modular, parallelizable, implemented in object-oriented Python, and publicly available. We have verified the images produced to be equivalent to those from traditional techniques to within a precision set by gridding coarseness. We have also validated our implementation on data observed with the Long Wavelength Array (LWA1). We provide a detailed framework for imaging with heterogeneous arrays and show that EPIC robustly estimates the input sky model for such arrays. Antenna layouts with dense filling factors consisting of a large number of antennas such as LWA, the Square Kilometre Array, Hydrogen Epoch of Reionization Array, and Canadian Hydrogen Intensity Mapping Experiment will gain significant computational advantage by deploying an optimized version of EPIC. The algorithm is a strong candidate for instruments targeting transient searches of Fast Radio Bursts (FRB) as well as planetary and exoplanetary phenomena due to the availability of high-speed calibrated time-domain images and low output bandwidth relative to visibility-based systems.
\end{abstract}



\section{Introduction}

Radio astronomy is entering an era in which interferometers of hundreds to thousands of individual antennas are needed to achieve desired survey speeds. Nowhere is this more apparent than at radio frequencies below 1.4 GHz. The study of the history of hydrogen gas throughout the universe's evolution is pushing technology development towards arrays of low-cost antennas with large fields of view and densely packed layouts. Similarly, the search for transient objects and regular monitoring of the time-dependent sky is driving instruments in the same direction with the added requirement of fast read-outs. A number of new telescopes around the world are based on this new paradigm, including the Hydrogen Epoch of Reionization Array\footnote{http://reionization.org} (HERA; \citealt{deb16}), the Murchison Widefield Array (MWA; \citealt{tin13,bow13}), the Donald~C.~Backer Precision Array for Probing the Epoch of Reionization (PAPER; \citealt{par10}), the LOw Frequency ARray (LOFAR; \citealt{van13}), the Canadian Hydrogen Intensity Mapping Experiment (CHIME; \citealt{ban14}), the Long Wavelength Array (LWA; \citealt{ell13}), and the low frequency Square Kilometer Array (SKA1-Low; \citealt{mel13}).

This paradigm shift requires a fundamentally new approach to the design of digital correlators \citep{lon00}. Modern correlators calculate the cross-power correlation between all antenna pairs in many narrow frequencies, forming \emph{visibilities}, the fundamental measurement of traditional radio interferometers. The computational requirements for a modern FX correlator scale with the number of antenna pairs, or the square of the number of antennas $\sim \Nant^2$ \citep{bun04}. For this reason traditional correlators have difficulty scaling to thousands of antennas. For example, the full HERA correlator for 352 dishes with 200 MHz of bandwidth requires 212 trillion complex multiply-accumulates per second (TMACS). Future arrays with thousands of collecting elements will require orders of magnitude more computation, making the correlator the dominant cost.

For certain classes of radio arrays there is an alternative to the FX correlator that can lower the computational burden by directly performing a spatial Fast Fourier Transform \citep[FFT;][]{coo65} on the electric fields measured by each antenna in the array at each time step, removing the cross-correlation step. This relieves the computational scaling from the harsh $\Nant^2$ to the more gentle envelope of $\sim\Ngrid\log_2\Ngrid$, where $\Ngrid$ is the number of grid points in the Fourier transform \citep[e.g.][]{mor11,teg09,teg10}. This architecture is often referred to as a ``direct imaging correlator'' because it eliminates the intermediate cross-correlation data products of the FX and XF correlators, but instead directly forms images from the electric field measurements.

Direct imaging correlators have begun to be explored on deployed arrays including the Basic Element for SKA Training II (BEST-2) array \citep{fos14}, the MIT EoR (MITEoR) experiment \citep{zhe14} using the {\it Omniscope} \citep{teg09,teg10}, and an earlier incarnation at higher frequencies with the intent of pulsar timing \citep{oto94,dai00}. However, each of these examples make assumptions about the redundancy of the array layout, and require that the collecting elements are identical. On the other hand, the MOFF algorithm achieves the same $\Ngrid \log_2 \Ngrid$ computational scaling without placing any restriction on antenna placement, can accommodate non-identical antennas, and is provably optimal \citep{mor11}. This algorithm uses the antenna beam patterns to grid the electric field measurements to a regular grid in the software holography/A-transpose fashion \citep{mor09,bha08,teg97b} before performing the spatial FFT. This process has been shown to mathematically produce a data product identical to images produced from traditional visibility-based techniques.

Here we present the first software implementation of the MOFF correlator, and announce the public release of the E-field Parallel Imaging Correlator (EPIC) code. EPIC is a highly parallel, object-oriented Python package that primarily implements the MOFF imaging algorithm, emulates real-life telescopes and FX/XF correlators in software, and includes a visibility-based imaging technique for reference. It is intended to provide a development platform to test different imaging approaches, characterize scaling relations and serve as a stepping stone for real-life GPU/FPGA-based implementation on telescopes.

We begin with a technical description of the algorithm in \S\ref{sec:math}, then discuss our particular implementation in \S\ref{sec:software}. We then verify the output data quality from our code in \S\ref{sec:verify} by presenting simulated images from both the EPIC correlator and comparing to a simulated FX correlator. We also demonstrate the performance with real-world data from the LWA1. In \S\ref{sec:versatility}, we present a mathematical framework for imaging with heterogeneous arrays and demonstrate the ability of EPIC to robustly image data from such arrays. In \S\ref{sec:analysis}, we explore the scalability of the algorithm in the context of several array design choices. We identify specific array design classes where the EPIC correlator is computationally more efficient, and in the field of transients, demands significantly lesser I/O bandwidth relative to visibility-based approaches. We conclude and discuss future research prospects in \S\ref{sec:conclusions}.

\section{Mathematical Framework}\label{sec:math}

We provide a brief summary of the mathematical equivalence of the MOFF and FX correlators detailed in \citet{mor11}. We first relate the image produced from visibilities to the electric fields of astrophysical sources, then show that operations can be reordered to produce the same images at a lower computational cost.

Electric fields from astrophysical sources, $\mathcal{E}(\hat{\mathbf{s}}, f, t)$, in the sky coordinate system denoted by the unit vector $\hat{\mathbf{s}}$, propagate towards the observer as:
\begin{align}
  E(\mathbf{r}, f, t) &= \int \mathcal{E}(\hat{\mathbf{s}},f,t)\,e^{-i 2\pi f\mathbf{r}\cdot\hat{\mathbf{s}}/c}\,\dif\Omega,
\end{align}
where, $i=\sqrt{-1}$, $\mathbf{r}$ denotes the observer's location, $f$ is the frequency of radiation, $c$ is the speed of light, $t$ denotes time, $\dif\Omega$ is the infinitesimal solid angle element on the celestial sphere whose normal is the unit vector $\hat{\mathbf{s}}$, and $E(\mathbf{r}, f, t)$ is the propagated electric field. Thus the propagated electric field is a linear superposition of the electric fields emanating from astronomical sources with appropriate complex phases. Ignoring wide-field effects, it can be simplified as a Fourier transform of the electric fields in the sky coordinates. 

An antenna, located at $\mathbf{r}_a$ (indexed by $a$), measures a phased sum of these propagated electric fields over its effective collecting area with an additive receiver noise:
\begin{align}\label{eqn:measured-E-field}
  E_a(f,t) &= \int W_a(\mathbf{r}-\mathbf{r}_a,f,t)\,E(\mathbf{r},f,t)\,\dif^2\mathbf{r} + n_a(f,t) \\
           &= \int W_a(\mathbf{r}-\mathbf{r}_a,f,t) \nonumber \\
           &\quad \times \left[ \int \mathcal{E}(\hat{\mathbf{s}},f,t)\,e^{-i 2\pi f\mathbf{r}\cdot\hat{\mathbf{s}}/c}\,\dif\Omega \right] \dif^2\mathbf{r} + n_a(f,t) \\
           &= \int \mathcal{W}_a(\hat{\mathbf{s}},f,t)\,\mathcal{E}(\hat{\mathbf{s}},f,t)\,e^{-i 2\pi f\mathbf{r}_a\!\cdot\,\hat{\mathbf{s}}/c}\,\dif\Omega + n_a(f,t)
\end{align}
where, $W_a(\mathbf{r},f,t)$ is the aperture electric field illumination pattern of the antenna and its Fourier transform, $\mathcal{W}_a(\hat{\mathbf{s}},f,t)$, is the directional antenna voltage response at a given frequency and time.

Interferometers measure {\it visibilities} -- the degree of coherence between electric fields measured by a pair of antennas \citep{van34,zer38,tho01}. A visibility, $V_{ab}$, can be written as:
\begin{align}
  V_{ab}(f,t) &= \left\langle E_a(f,t)E_b^\star(f,t) \right\rangle_t \label{eqn:cc-vis}\hfill\\
              &= \left\langle \left[ \int \mathcal{W}_a(\hat{\mathbf{s}},f,t)\,\mathcal{E}(\hat{\mathbf{s}},f,t)\,e^{-i 2\pi f\mathbf{r}_a\!\cdot\,\hat{\mathbf{s}}/c}\,\dif\Omega\right.\right. \nonumber\\
              &\qquad\qquad\qquad\qquad\qquad\qquad\qquad\quad + \left. n_a(f,t) \vphantom{\int}\right] \nonumber\\
              &\quad \times \left[ \int \mathcal{W}^\star_b(\hat{\mathbf{s'}},f,t)\,\mathcal{E}^\star (\hat{\mathbf{s'}},f,t)\,e^{i 2\pi f\mathbf{r}_b\!\cdot\,\hat{\mathbf{s'}}/c}\,\dif\Omega^\prime \right. \nonumber\\
              &\qquad\qquad\qquad\qquad\qquad\qquad\qquad\quad + \left.\left. n^\star_b(f,t) \vphantom{\int}\right] \right\rangle_t \nonumber\\
              &= \iint \mathcal{W}_a(\hat{\mathbf{s}},f,t) \mathcal{W}^\star_b(\hat{\mathbf{s'}},f,t) \left\langle \mathcal{E}(\hat{\mathbf{s}},f,t) \mathcal{E}^\star(\hat{\mathbf{s'}},f,t) \right\rangle_t \nonumber\\
              &\qquad\qquad\qquad\qquad \times e^{-i 2\pi f(\mathbf{r}_a\!\cdot\,\hat{\mathbf{s}}-\mathbf{r}_b\!\cdot\,\hat{\mathbf{s'}})/c}\,\dif\Omega\,\dif\Omega^\prime \nonumber\\
              &\qquad +\textrm{noise-like cross-terms.}
\end{align}
where we have brought the time average into the integral under the assumption that the aperture illumination pattern does not change over the time-scale of the averaging and $\star$ denotes complex conjugation. The noise-like cross-terms ideally have zero mean. This expression can be further simplified using the sky brightness, $\mathcal{I}(\hat{\mathbf{s}},f,t)$, as $\left\langle \mathcal{E}(\hat{\mathbf{s}},f,t)\,\mathcal{E}^\star(\hat{\mathbf{s'}},f,t) \right\rangle_t = \mathcal{I}(\hat{\mathbf{s}},f,t)\,\delta(\hat{\mathbf{s}}-\hat{\mathbf{s'}})$, and defining the antenna-pair sky power response function (or the directional antenna-pair power pattern), $\mathcal{B}_{ab}(\hat{\mathbf{s}})~\equiv~\mathcal{W}_a(\hat{\mathbf{s}},f,t)~\mathcal{W}^\star_b(\hat{\mathbf{s}},f,t)$. The result is the visibility expressed in terms of the sky brightness, the power pattern, and uncorrelated noise terms which we group into $n_{ab}(f,t)$.
\begin{align}\label{eqn:measurement-eqn-1}
  V_{ab}(f,t) &= \int \mathcal{B}_{ab}(\hat{\mathbf{s}},f,t)\,\mathcal{I}(\hat{\mathbf{s}},f,t)\,e^{-i 2\pi f\mathbf{r}_{ab}\!\cdot\,\hat{\mathbf{s}}/c}\,\dif\Omega \nonumber\\
  &\qquad\qquad + n_{ab}(f,t),
\end{align}
where, the baseline coordinate $\mathbf{r}_{ab}\equiv\mathbf{r}_a-\mathbf{r}_b$ is the vector separation between the two antennas. This signifies that the visibility ($V_{ab}$) measured between a pair of antennas is obtained by multiplying the sky brightness $\mathcal{I}(\hat{\mathbf{s}},f,t)$ by the antenna power response $\mathcal{B}_{ab}(\hat{\mathbf{s}},f,t)$ and Fourier-transforming from the directional coordinates ($\hat{\mathbf{s}}$) to the measurement plane, which are then sampled at the locations of the antenna spacings (or baselines), namely, $\mathbf{r}_{ab}$, and added to the noise $n_{ab}$. 

This can be equivalently re-written as:
\begin{align}\label{eqn:software-holography}
  V_{ab}(f,t) &= \int B_{ab}(\mathbf{r}_{ab}-\mathbf{r},f,t) \nonumber\\ 
              &\qquad \times \left[\int I(\hat{\mathbf{s}},f,t)\,e^{-i 2\pi f\mathbf{r}.\hat{\mathbf{s}}/c}\,\dif\Omega\right]\dif^2\mathbf{r}\,+\, n_{ab}(f,t),
\end{align}
where, $B_{ab}(\mathbf{r},f,t)$ denotes the power response of the antenna pair obtained in the measurement plane by a spatial Fourier transform of $\mathcal{B}_{ab}(\hat{\mathbf{s}},f,t)$. Effectively, the multiplication in image space by $\mathcal{B}_{ab}(\hat{\mathbf{s}},f,t)$ has been replaced by a convolution with $B_{ab}(\mathbf{r},f,t)$ in the measurement plane. This is the software holographic equivalent of a traditional FX correlator output. For context, the term ``holography'' is derived from holographic measurements of the complex antenna aperture illumination pattern \citep{ben76,sco77}. In the present context, ``software holography'' refers to the application of this antenna aperture illumination pattern in software for imaging.

Hereafter, we adopt the matrix notation of \citet{mor11}, where vectors are represented with single coordinates, and matrices are represented by two coordinates denoting the spaces the operator transforms between. Since each frequency is processed independently, we drop the dependence on $f$. In this notation, the above measurement equation can be expressed as:
\begin{align}
  \mathbf{m}(\mathbf{v}) &= \mathbf{B}(\mathbf{v},\mathbf{u})\,\mathbf{F}(\mathbf{u},\hat{\mathbf{s}})\,\boldsymbol{\mathcal{I}}(\hat{\mathbf{s}}) + \mathbf{n}(\mathbf{v}),
\end{align}
where the sky brightness $\boldsymbol{\mathcal{I}}(\hat{\mathbf{s}})$ is Fourier-transformed using $\mathbf{F}(\mathbf{u},\hat{\mathbf{s}})$ and the resultant spatial coherence function is weighted and summed using the antenna power response, $\mathbf{B}(\mathbf{v},\mathbf{u})$ in $uv$-space sampled at the baseline location to obtain the measured visibilities:
\begin{align}
  \mathbf{m}(\mathbf{v}) &= \left\langle \mathbf{E}(\mathbf{a})\,\mathbf{E}^\star(\mathbf{a}^\prime)\right\rangle_t, \label{eqn:matrix-cc-vis}
\end{align}
where $\mathbf{m}(\mathbf{v})$ denotes visibilities measured by cross-correlating measured antenna electric fields over all possible pairs of $\mathbf{a}$ and $\mathbf{a}^\prime$. It is the same as equation~\ref{eqn:cc-vis} written in matrix notation.

Using the optimal map-making formalism \citep{teg97a,teg97b}, a software holography image is formed using \citep{mor09}:
\begin{align}
  \boldsymbol{\mathcal{I}}^\prime(\hat{\mathbf{s}}) &= \mathbf{F}^\textrm{T}(\hat{\mathbf{s}},\mathbf{u})\,\mathbf{B}^{\,\textrm{T}}(\mathbf{u},\mathbf{v})\,\mathbf{N}^{-1}(\mathbf{v},\mathbf{v})\,\mathbf{m}(\mathbf{v}) \label{eqn:dirty-image-FX}
\end{align}
where the measured visibilities are weighted by the inverse of the system noise, followed by a gridding process using the holographic antenna power response as the gridding kernel, followed by a Fourier transform to create an image $\boldsymbol{\mathcal{I}}^\prime(\hat{\mathbf{s}})$. This is the optimally weighted estimate of the true image $\boldsymbol{\mathcal{I}}(\hat{\mathbf{s}})$ given the visibility measurements.

The intermediate step of gridding with the antenna power response can be expressed as a convolution of a data vector generated by gridding the electric fields directly with the antenna illumination pattern.
\begin{align}
\mathbf{B}^{\,\textrm{T}}(\mathbf{u},\mathbf{v})\,\mathbf{N}^{-1}(\mathbf{v} ,\mathbf{v})\,\mathbf{m}(\mathbf{v}) &= \left\langle \left[\mathbf{W}_a(\mathbf{r},\mathbf{a})\,\widetilde{\mathbf{N}}^\textrm{T}\!(\mathbf{a},\mathbf{a})\, \mathbf{E}^\star(\mathbf{a})\right]\right. \nonumber\\ 
&\,\quad\ast\left.\left[\mathbf{W}^\textrm{T}_a(\mathbf{r},\mathbf{a})\,\widetilde{\mathbf{N}}\!(\mathbf{a},\mathbf{a})\,\mathbf{E}(\mathbf{a})\right]\right\rangle_t, \label{eqn:e-field-conv}
\end{align}
where, $\mathbf{N}^{-1}$ has been expressed as $\widetilde{\mathbf{N}}^\textrm{T}\widetilde{\mathbf{N}}$.

We can then use the multiplication-convolution theorem to move the convolution in Equation~\ref{eqn:e-field-conv} to a product after the Fourier transform in equation~\ref{eqn:dirty-image-FX}.
\begin{align}
  \boldsymbol{\mathcal{I}}^\prime(\hat{\mathbf{s}}) &= \left\langle \left|\,\mathbf{F}^\textrm{T}(\hat{\mathbf{s}},\mathbf{r})\,\mathbf{W}^\textrm{T}(\mathbf{r},\mathbf{a})\,\widetilde{\mathbf{N}}(\mathbf{a},\mathbf{a})\,\mathbf{E}(\mathbf{a})\,\right|^2\right\rangle_t. \label{eqn:dirty-image-MOFF}
\end{align}
The term inside the angular brackets before squaring has a very similar form as that in equation~\ref{eqn:dirty-image-FX}. It signifies that the measured antenna electric fields are weighted by the antenna noise, weighted and gridded by the antenna aperture kernel, Fourier-transformed and finally squared to obtain the same image estimate that would have been obtained using equation~\ref{eqn:dirty-image-FX}. 

Equation~\ref{eqn:dirty-image-MOFF} is the optimal imaging equation used by the MOFF algorithm. While mathematically equivalent to equation~\ref{eqn:dirty-image-FX}, squaring in image space rather than convolving in $uv$ space potentially saves orders of magnitude in computation.

There are some important differences between the two techniques:
\begin{enumerate}
\item The time-averaging cannot be performed on a stochastic measurement but only on its statistical properties. In visibility-based imaging, the visibilities measured between antenna pairs represent spatial correlations which can be time-averaged followed by gridding and imaging. However, in MOFF imaging both antenna and gridded electric fields are stochastic and therefore must be imaged and squared before time-averaging. 
\item In visibility-based imaging, electric fields measured by antennas are not correlated with themselves and hence lack zero spacing measurements. In contrast, in MOFF imaging, since the gridded electric fields are imaged and squared, they retain information from auto-correlated electric fields at zero spacing. However, we will show in \S\ref{sec:rm-autocorr} that they can be removed by fitting and subtracting the zero-spacing contribution during imaging.
\end{enumerate} 

\section{Software Implementation}\label{sec:software}

We have implemented the MOFF imaging technique in our ``E-field Parallel Imaging Correlator'' -- a highly parallelized Object Oriented Python package,\footnote{The E-field Parallel Imaging Correlator (EPIC) package can be accessed at https://github.com/nithyanandan/EPIC} now publicly available. Besides implementing the MOFF imaging algorithm it also includes a simulator for generating electric fields from a sky model and a visibility-based imaging routine using the software holography technique.

EPIC can accept dual-polarization inputs and produce images of all four instrumental cross-polarizations. Currently, two data input formats exist for reading in the electric field time samples measured by the antennas -- simulated electric fields based on a sky model using the simulator packaged with EPIC; and LWA1 data. Efforts to build interfaces for data from other telescopes are underway.

\begin{figure}
  \includegraphics[width=\columnwidth]{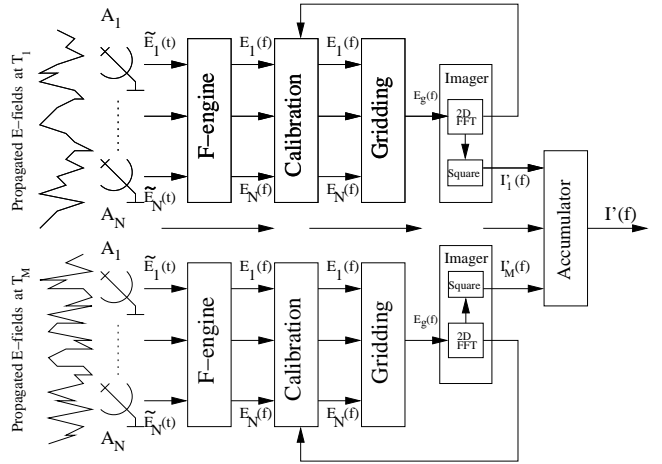}
  \caption{Flowchart of MOFF imaging in EPIC. The propagated electric fields shown on the left are measured as time-series $\widetilde{E}_1(t)\ldots \widetilde{E}_\textrm{N}(t)$ by the antennas $\textrm{A}_1\ldots \textrm{A}_\textrm{N}$ which are then Fourier-transformed by the F-engine to produce electric field spectra $E_1(f)\ldots E_\textrm{N}(f)$. They are calibrated and gridded. The gridded electric fields $E_\textrm{g}(f)$ from each time series are imaged to produce images $\mathcal{I}^\prime_1(f)\ldots \mathcal{I}^\prime_\textrm{M}(f)$. These images are time-averaged to obtain the final image $\mathcal{I}^\prime(f)$.}
  \label{fig:MOFF-flowchart}
\end{figure}

Fig.~\ref{fig:MOFF-flowchart} shows the flowchart for MOFF imaging. The propagated electric fields are shown on the left at different time stamps, $t_1\ldots t_\textrm{M}$. At each time stamp, the electric fields measured by antennas are denoted by $\widetilde{E}_1(t)\ldots \widetilde{E}_\textrm{N}(t)$. The F-engine performs a temporal Fourier transform on the electric field time-series to obtain electric field spectra $E_1(f)\ldots E_\textrm{N}(f)$ ($\mathbf{E}(\mathbf{a})$ in matrix notation) for each of the antennas. Each of the complex antenna gains are calibrated to correct the corresponding electric field spectra. These calibrated electric fields are gridded using an antenna-based gridding convolution function after which it is spatially Fourier-transformed and squared to obtain an image cube for every time stamp. These images are then time-averaged to obtain the accumulated image $\mathcal{I}^\prime(f)$ ($\boldsymbol{\mathcal{I}^\prime}(\hat{\mathbf{s}})$ in matrix notation).

\begin{figure}
  \includegraphics[width=\columnwidth]{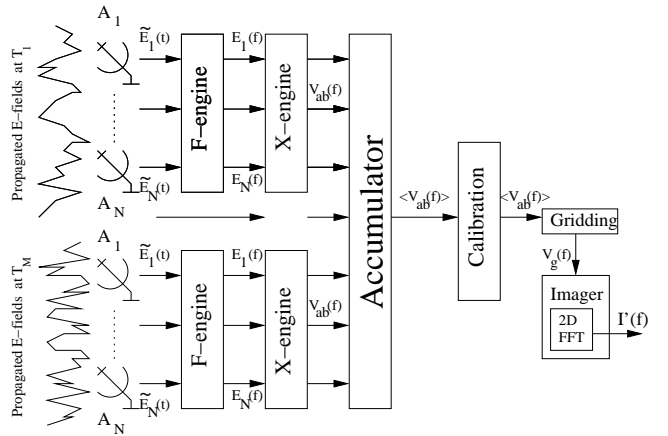}
  \caption{Flowchart of visibility-based software holographic imaging in EPIC. The FX process flow shares the F-engine with the MOFF process. Following the F-engine, the electric fields pass through the X-engine to produce visibilities $V_\textrm{ab}(f,t)$ which are calibrated and time-averaged. Then they are gridded to obtain the gridded visibilities $V_\textrm{g}(f)$ which are then Fourier-transformed to obtain the image, $\mathcal{I}^\prime(f)$.}
  \label{fig:FX-flowchart}
\end{figure}

Fig.~\ref{fig:FX-flowchart} shows the flowchart for a visibility-based software holographic imaging from a FX correlator. The antenna-based F-engine is identical to that in the MOFF processing. The electric field spectra from each antenna are then cross-multiplied in the X-engine with those from all other antennas to obtain the visibilities $V_\textrm{ab}(f,t)$ ($\mathbf{m}(\mathbf{v})$ in matrix notation). They are calibrated and time-averaged to obtain $\langle V_\textrm{ab}(f)\rangle$ which are then gridded and imaged to obtain the image $\mathcal{I}^\prime(f)$. The $\mathcal{I}^\prime(f)$ obtained from both techniques are mathematically identical as explained in \S\ref{sec:math}.

A high level software architecture of the EPIC package is described in the appendix~\ref{sec:software-modules} for the interested reader. Here we discuss the components of these architectures in detail. 

\par\medskip
\noindent {\bf Antenna-to-Grid Mapping}
\par\medskip
\noindent A grid is generated on the coordinate system in which antenna locations are specified. The grid spacing can be controlled by the user. By default, it is set to be $\le\lambda/2$ even at the lowest wavelength to ensure there is no aliasing even from regions of the sky far away from the field of view. The number of locations on the grid is restricted to be a power of 2 for efficient use of FFT. 

The gridding kernel in the simplest case is given by the antenna aperture illumination function, $W(\mathbf{r}-\mathbf{r}_a)$, which can be specified either by a functional form or as a table of values against locations around the antennas. A nearest neighbour mapping from all antenna footprints to grid locations is created using an efficient k-d tree algorithm \citep{man99}. There is no restriction here that the aperture illumination function has to be identical across antennas. 

\begin{figure}
  \includegraphics[width=\columnwidth]{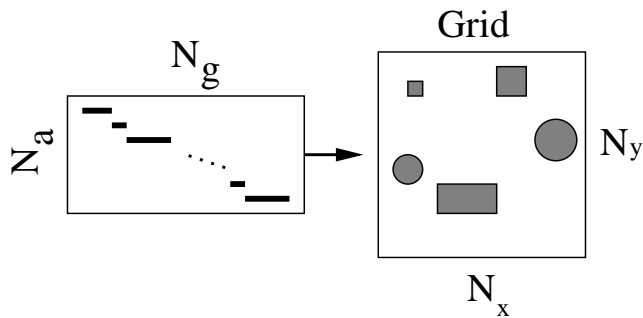}
  \caption{Block diagram of an antenna-to-grid mapping. A sparse block-diagonal     matrix of total size $\Ngrid\times \Nant$ is created where each block contains roughly the number of pixels covered by the respective kernel. The antenna aperture illumination kernels do not have to be identical to each other. A discrete set of arbitrarily placed antennas are now placed on to a regular grid.}
  \label{fig:a2g-mapping}
\end{figure}

In the most general case, this gridding kernel could contain information on $w$-projection effects, and other time-dependent ionospheric effects. For a stationary antenna array in the absence of any time-dependent effects, this mapping must only be determined once in the antenna array coordinate frame. The antenna-to-grid mapping matrix, $\mathbf{M}(\mathbf{r},\mathbf{a})$ is described as a transformation matrix from the space of measured electric fields by the antennas ($\mathbf{a}$) to the antenna array grid denoted by the coordinate $\mathbf{r}$. Since each antenna occupies a footprint typically the size of its aperture, $\mathbf{M}(\mathbf{r},\mathbf{a})$, which is generally of size $\Ngrid\times \Nant$, reduces to a sparse block-diagonal matrix with only $\Nant$ blocks and roughly $N_\textrm{k}$ non-zero entries per block, where $N_\textrm{k}$ is the number of grid points that fall inside an antenna's footprint. This sparse matrix is stored in a Compressed Sparse Row (CSR) format. Fig.~\ref{fig:a2g-mapping} illustrates the antenna-to-grid mapping matrix and the grid containing the mapped aperture footprints of the antennas.

\par\medskip
\noindent {\bf Temporal Fourier transform}
\par\medskip
\noindent This module is common to the MOFF and visibility-based imaging techniques. Time samples of electric fields measured by the antenna and digitized by the A/D converter is Fourier-transformed to generate electric field spectra. This step can be parallelized across antennas. The output is then fed to either MOFF or visibility-based imaging pipelines.

\par\medskip
\noindent {\bf Calibration}
\par\medskip
\noindent Calibration of direct imaging correlators remains a challenge. Contrary to the FX data flow, direct imagers mix the signals from all antennas before averaging and writing to disk. It is therefore essential to apply gain solutions before the gridding step. Previous efforts have resorted to applying FX-generated calibration solutions \citep{zhe14,fos14}, or integrating a dedicated FX correlator which periodically forms the full visibility matrix \citep{wij09,dev09}. 

In a companion paper \citep{bea16}, we demonstrate a novel calibration technique (EPICal) which leverages the data products formed by direct imaging correlators to estimate antenna complex gains. This method correlates the antenna electric field signals with an image pixel from the output of the correlator in the feedback calibration fashion outlined in \citealt{mor11} (illustrated in Fig.~\ref{fig:MOFF-flowchart} by the arrow leading from the imager to the calibration block). Furthermore, it allows for arbitrarily complex sky models, places no restrictions on array layout, and accounts for non-identical antenna beam patterns. Direction-dependent calibration can be achieved by correlating antenna signals with output pixels in the direction of $N_\textrm{c}$ calibration sources, then fitting for a functional model of the sky. Since antennas are only correlated with calibrator pixels, the computational complexity scales as $\sim N_\textrm{a} N_\textrm{c}$. The calibration performance of EPIC is explored for varying levels of completeness of sky model in \citet{bea16}.

The calibration module included in EPIC allows for application of pre-determined calibration solutions, or can solve for the complex gains using the EPICal algorithm.

\par\medskip
\noindent {\bf Gridding Convolution}
\par\medskip
\noindent The antenna array aperture illumination over the entire grid, $W(\mathbf{r})$, is obtained by a projection of the individual antenna aperture illuminations:
\begin{align}\label{eqn:gridding-convolution}
  W(\mathbf{r}) &= \sum_a W_a(\mathbf{r}-\mathbf{r}_a),
\end{align}
or in matrix notation,
\begin{align}
  \mathbf{W}(\mathbf{r}) &= \mathbf{M}(\mathbf{r},\mathbf{a})\,\mathbf{I}(\mathbf{a}),
\end{align}
where, $\mathbf{I}(\mathbf{a})$ is a row of ones denoting coverage of the grid by kernel footprints of antennas. This is achieved by efficient multiplication with the sparse matrix created in the antenna-to-grid mapping process using the sparse matrix module in Python SciPy package. Unless $\mathbf{W}(\mathbf{r})$ includes time-dependent effects of the ionosphere or the instrument, it needs to be computed just once for the entire observation. However, the gridding of electric fields must be computed at every readout of the electric field spectra,
\begin{align}
  \mathbf{E}(\mathbf{r}) &= \mathbf{M}(\mathbf{r},\mathbf{a})\,\mathbf{E}(\mathbf{a}).
\end{align}

\par\medskip
\noindent {\bf Spatial Fourier Transform}
\par\medskip
\noindent Before the spatial Fourier transform, the gridded electric fields are padded with zeros in order to match the grid size and angular size of each image pixel that would have been obtained with the software holography output from an FX correlator. 

In MOFF imaging, these are spatially Fourier-transformed followed by squaring at every time stamp for every frequency channel. In visibility-based imaging, the spatial Fourier transform is performed only once per integration time-scale and does not include a squaring operation.

\par\medskip
\noindent {\bf Time-averaging}
\par\medskip
\noindent In MOFF imaging, the measured antenna electric fields and the corresponding holographic electric field images are zero-mean stochastic quantities. Hence, they cannot be time-averaged to reduce noise. The statistical quantity stable with time in this case is the square of the holographic electric field image. Thus, squared images have to be formed at every instant of time before averaging as indicated in equation~\ref{eqn:dirty-image-MOFF}.

In contrast, visibilities measured by an antenna are statistically stable within an integration time interval. Hence, they are averaged after (and sometimes before) calibration as shown in equation~\ref{eqn:cc-vis}. It is advantageous to average them in visibilities before imaging because the repeated cost of spatial FFT can be avoided. Since this averaging has been performed already on the visibilities over an integration time-scale, the imaging step has to be performed only once per integration cycle. 

\section{Verification}\label{sec:verify}

In order to verify the accuracy of the EPIC code, we characterize the images produced through simulations. We simulate electric field streams from a model sky and process the data through both the MOFF and a visibility-based imaging algorithm. We then compare the output images to demonstrate their equivalence.

\subsection{Simulations}\label{sec:sim}

We use the EPIC simulator to generate stochastic electric field samples from a sky model consisting of 10 point sources of random flux densities $>10$~Jy each at random locations. In our simulations, we use 64 frequency channels each of width $\Delta f = 40$~kHz. The number of time stamps integrated in one integration cycle was kept at eight where the duration of each time-series is $1/\Delta f=25\,\mu$s long. We use the MWA array layout \citep{bea12} for demonstration. Only the inner 51 tiles within a square bounding box of 150~m on each side were used. We assumed all tiles are identical and have a square-shaped electric field illumination footprint 4.4~m on each side. Besides the stochastic sky noise present in the simulated electric fields, no noise from the instrument is added.

\subsection{Antenna auto-correlations}\label{sec:rm-autocorr}

Before the outputs can be compared, we describe the elimination of a distinct difference between the two techniques. The squaring operation under MOFF imaging in the image plane introduces antenna auto-correlations around the zero-spacing in the $uv$-plane which are absent in traditional visibility-based imaging. In order to facilitate a robust comparison between MOFF and visibility-based imaging techniques, these auto-correlations are removed from the MOFF algorithm output, which is otherwise not an essential part of the core algorithm. We describe below how they are removed. 

\begin{figure}
  \includegraphics[width=\columnwidth]{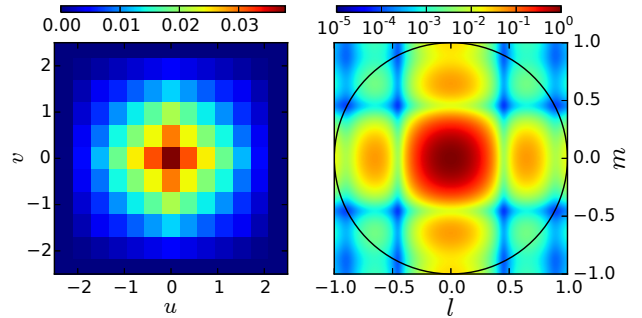}
  \caption{Auto-correlation of weights of a square shaped antenna aperture in the $uv$ plane (left) and the corresponding directional antenna power response on the sky (right) in coordinates specified by direction cosines. The antenna auto-correlation weights are normalized to a sum of unity yielding a peak response of unity in the antenna's directional power pattern on the sky. The colour scale for the directional power pattern is logarithmic. The black circle indicates the sky horizon and values beyond it are not physical and hence ignored.}
  \label{fig:autocorr_wts_PB}
\end{figure}

The shape and extent of these auto-correlations can be estimated from the antenna aperture illumination pattern. The aperture illumination patterns are already available from the gridding step. Fig.~\ref{fig:autocorr_wts_PB} shows the estimated weights from antenna auto-correlations in the $uv$-plane (left) and the corresponding response in the image plane (right). The latter is simply the directional antenna power response. 

We perform the inverse Fourier transform of the squared images and beams back to the $uv$ plane and subtract the estimated auto-correlation kernel scaled to the peak value centred at the zero-spacing pixel. The final averaged image is obtained by Fourier-transforming the $uv$ plane data and weights with the auto-correlations subtracted to the image plane. These images are now comparable to those obtained from visibility-based imaging. This step of removing auto-correlations needs to be performed only once per integration time-scale and does not add significant cost to the full operation.

\subsection{Comparison of outputs}\label{sec:diff}

We investigate the two imaging algorithms for differences from the point of view of the quality of their outputs. We begin by comparing the images produced with the two approaches. 

\begin{figure}
  \includegraphics[width=\columnwidth]{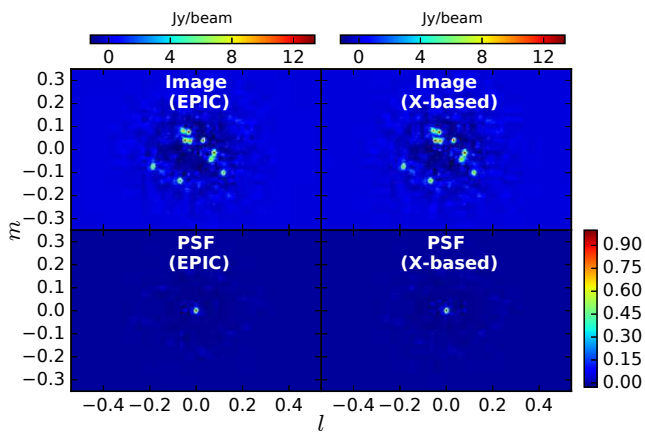}
  \caption{Weighted dirty images (top) and synthesized beams (bottom) obtained from simulated data using EPIC implementation of antenna-based MOFF algorithm (left) and visibility-based imaging (right). The antenna auto-correlations at zero-spacing have been removed from the MOFF images. The images in either case reconstruct the sources at the right locations with the fluxes attenuated expected after multiplication by the antenna power pattern squared. The synthesized beams from the two algorithms are well matched in size and shape. The overall modulation by the power pattern is seen clearly in both images.}
  \label{fig:MOFF-FX-image}
\end{figure}

Fig.~\ref{fig:MOFF-FX-image} shows the weighted dirty images (top) and synthesized beams (bottom) obtained with antenna-based MOFF and FX visibility-based imaging algorithms packaged in EPIC. The antenna auto-correlations that correspond to zero-spacing have been removed from the correlated weights and data in the $uv$ plane, the MOFF image and the corresponding synthesized beam as described in \S\ref{sec:rm-autocorr}. The reconstructed sky image has the simulated sources at the expected sky positions in either case. Both algorithms result in images and synthesized beams that are well matched with each other. As expected, their fluxes are attenuated by a weighting proportional to the square of the antenna power pattern corresponding to a uniform square aperture.

\begin{figure}
  \includegraphics[width=\columnwidth]{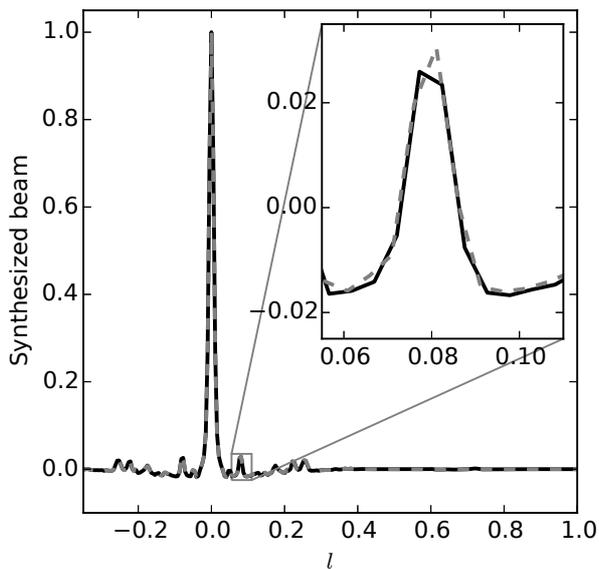}
  \caption{Synthesized beams weighted by antenna power pattern sliced at $m=0$ obtained with MOFF algorithm using EPIC (solid black line) and visibility-based imaging (dashed grey line). The two appear almost identical. The inset provides a magnified view of differences in the synthesized beam slices between the two techniques at levels $\lesssim 0.5$\% relative to the peak. These are attributed to differences that arise in gridding and depends on the coarseness of the grid.}
  \label{fig:MOFF-FX-psf-diff}
\end{figure}

We examine in detail the respective synthesized beams in each case in Fig.~\ref{fig:MOFF-FX-psf-diff}. Slices at $m=0$ of the synthesized beams weighted by the antenna power pattern are shown for MOFF method using EPIC (solid black) and visibility-based (dashed grey) imaging. The two are found to match well. A magnified view shows that some differences at the level of $<0.5$\% exist in some regions. These are attributed to differences in the $uv$-plane antenna cross-correlation weights in the two methods which in turn arises due to the amount of coarseness in grid spacing as described below.

\begin{figure}
  \includegraphics[width=\columnwidth]{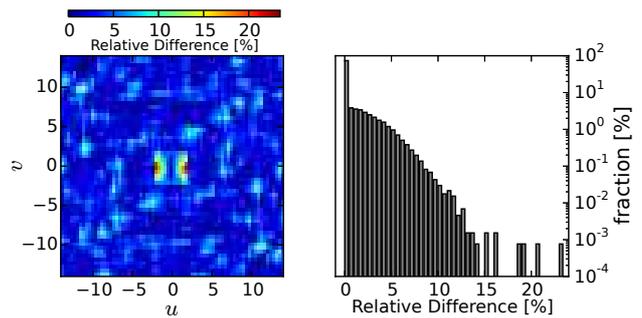}
  \caption{{\it Left}: Differences (in \%) in cross-correlated weights in the $uv$ plane relative to the peak. The biggest difference ($\sim 20$\%) is found around the zero-spacing corresponding to auto-correlations of antenna weights due to gridding differences augmented by the relatively high number density of antenna auto-correlation footprints in that region. In most of the other regions, differences of the order of less than a few percent are seen. {\it Right}: Histogram (expressed as percentage) of the relative differences (in \%) between antenna cross-correlation weights shown on the left binned in intervals of 0.5\%. Over 70\% of the $uv$-cells do not differ by more than 0.5\% and over 90\% of $uv$-cells only differ by $<5$\% for the gridding coarseness used.} 
  \label{fig:MOFF-FX-uvwts-diff}
\end{figure}

The left panel of Fig.~\ref{fig:MOFF-FX-uvwts-diff} shows differences (in percentage relative to peak) in cross-correlated weights obtained with MOFF imaging in EPIC and visibility-based imaging. The maximum difference appears near the centre of the $uv$-plane corresponding to antenna auto-correlations which have not been perfectly removed in the former. In other regions of the $uv$-plane, the differences are of the order of a few percent. The reason for these differences is described below.

The gridding step in MOFF imaging samples the antenna footprint (either in analytic or lookup table formats) at the grid locations. Coverage of grid pixels by an antenna footprint may be $\sim 1$ pixel narrower particularly at the edge of the footprint along one or both directions relative to that from another identical antenna but with a fractionally different location relative to the grid. This depends on the exact location of the centre of the antenna relative to the grid and the coarseness of grid spacing. This first order loss of precision of the sampled footprint propagates to second order ($\sim 2$ pixels) upon correlation of the discretized weights. In other words, the correlated weights may suffer further loss of precision in their sampled footprint after correlation of two footprints each of which could be less precise to first order. On the other hand, in visibility-based imaging, a directly sampled $uv$ plane antenna power response (which is mathematically identical to the correlation of individual antenna footprints) centred on a baseline has a loss in precision at most to first order. Thus, although in the limit of infinitesimally small grid spacing they should be identical, the coarseness of grid spacing introduces subtle differences between the two.

These differences which are dependent on the coarseness of grid spacing can be mitigated by making the grid spacing finer at the expense of increased computational cost. Residuals centred around zero-spacing can also be lowered by subtracting each auto-correlation of antenna weights separately by using the shape and extent of the sampled footprint appropriate for that specific antenna aperture. This is a general solution applicable even in case of heterogeneous antenna arrays and is under active development for EPIC.

\begin{figure}
  \includegraphics[width=\columnwidth]{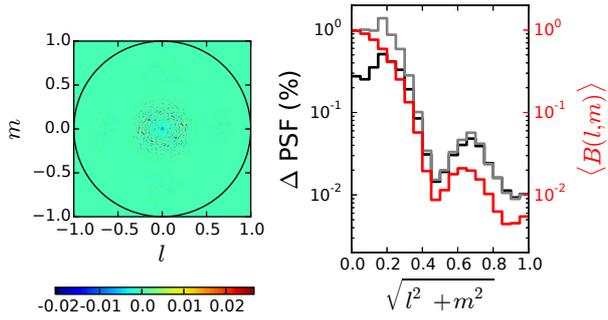}
  \caption{Map of difference between the synthesized beams obtained with the two methods (left) and radial statistics of the synthesized beams and their differences (right). The radial variations of rms of the synthesized beam (grey) and that of the rms of the difference (black) are shown as percentage of the peak synthesized beam. Radially averaged directional antenna power response in absolute scale is shown in red and is to be read with the scale on the right side of the axis. The maximum difference is of the order of a few percent. Amplitude of the rms of the difference is modulated by the power pattern of the antenna.}
  \label{fig:image-psf-diff}
\end{figure}

We study the effect of the differences in gridded weights on the image plane. Fig.~\ref{fig:image-psf-diff} shows the difference between the synthesized beams obtained with the two methods. A difference map between the two synthesized beams is shown on the left. The amplitude of the difference appears to be modulated by the directional power response of the antenna. On the right, in radial bins, the rms of the synthesized beam (grey) and the rms of the difference map (black) are plotted in percentage units relative to the peak (to be read using the axis on the left side of the plot). The antenna power pattern (red; to be read using the scale on the right) is plotted for reference. 

The synthesized beam rms is proportional to the antenna power pattern as expected from a point spread function uncorrected for the antenna power pattern. The rms of differenced synthesized beams is also modulated by the antenna power pattern. The rms of the difference is definitely lesser than the rms of the synthesized beam in the central regions up to $(l^2+m^2)^{1/2}\lesssim 0.3$. This implies that the beams are well matched in the central regions. In the outer regions, their mismatch is comparable to the rms of synthesized beams. These findings are consistent with Fig.~\ref{fig:MOFF-FX-psf-diff}. This indicates the two synthesized beams are not randomly different from each other in which case the rms of the difference would have been $\approx \sqrt{2}$ higher than the rms of the each of the synthesized beams. This means that while differences exist, large fractions of them are still well matched to each other even out to the horizon. Thus the rounding errors in gridding do not affect the statistics of the images or the synthesized beams.

\subsection{Application to LWA1 data}\label{sec:LWA-data}

Here we demonstrate our software using narrow band data from the LWA1 station in New Mexico. This data is in LWA1 narrow-band transient buffer (TBN) format from 255 antennas within roughly a diameter of 100~m. The data is centred at a frequency of 74.03~MHz, with a sample rate (equal to the bandwidth) of 100~kHz with 512 complex time samples per antenna in a A/D writeout time-scale of 5.12~ms, a frequency resolution of 195.3125~Hz and dual polarization. There are 391 such writeouts (or time stamps; each contains 512 time samples at 100~kHz sampling) yielding a total duration of 2~s. 

We corrected the cable delays, but otherwise assume the data is sufficiently calibrated to image directly. A detailed demonstration of EPICal on this data is presented in \citet{bea16}.

\begin{figure}
  \includegraphics[width=\columnwidth]{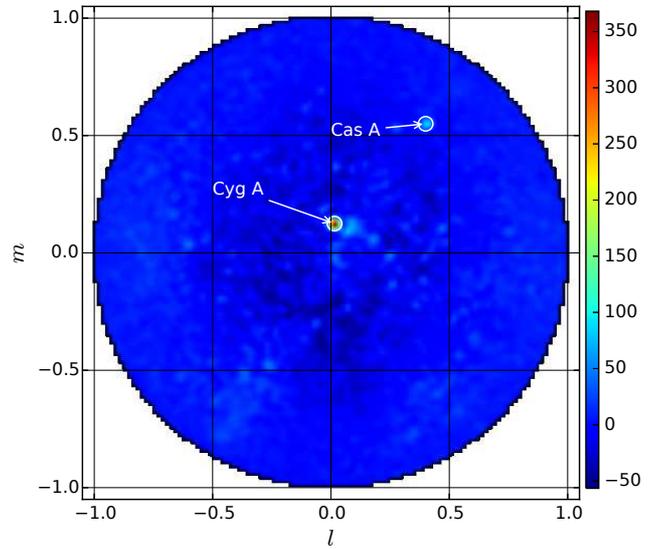}
  \caption{Image from LWA1 TBN data obtained with MOFF imaging using EPIC package after averaging over 20~ms and $\approx 80$~kHz. The x- and y-axes denote direction cosines $l$ and $m$ respectively. The antenna voltages are compensated for their respective delays. The flux scale is arbitrary. Locations of Cyg~A and Cas~A are annotated.}
  \label{fig:LWA-image}
\end{figure}

Fig.~\ref{fig:LWA-image} shows the image produced with MOFF imaging packaged in EPIC after averaging over $\approx$~20~ms (four writeouts) of data and the inner $\approx 80$\% of bandwidth (roughly 80~kHz). The image is shown in direction cosine coordinates -- $l$ along the horizontal axis and $m$ along the vertical axis. The flux scale is arbitrary. Even in this proof-of-concept demonstration, we see Cyg~A and Cas~A prominently as annotated, thus validating the functionality of EPIC.

\section{Imaging with Heterogeneous Antenna Arrays}\label{sec:versatility}

One of the advanced features of the MOFF algorithm and its implementation via EPIC is the ability to naturally account for heterogeneity of antennas while still producing images with maximal information \citep{mor11}. Unlike the assumption frequently made in existing direct imaging techniques that all antennas are identical \citep{oto94,dai00,teg09,teg10,fos14,zhe14}, EPIC does not assume $W_a(\mathbf{r}-\mathbf{r}_a)$ or $B_{ab}(\mathbf{r}-\mathbf{r}_{ab})$ for different antennas are identical. However, any {\it a priori} knowledge available that certain sets of antennas have identical illumination patterns can be easily passed on to EPIC to avoid redundant computations and thus increase efficiency. Here we present key highlights from the generic methodology detailed in appendix~\ref{sec:math-versatility} to understand the effective angular weighting in the image obtained with EPIC from a heterogeneous array. 

Heterogeneity in antenna arrays occurs primarily in two scenarios. The first is when the array is intended to be identical but differences in aperture and electrical properties arise due to deviations during the manufacturing process or from random deviant behaviour such as differences in gains between dipoles in a phased array. The second scenario is when the array is designed to be heterogeneous to gain specific advantages such as sampling a wider range of spatial frequencies and decoupling the sky brightness distribution from the antenna power pattern. This has motivated a number of heterogeneous arrays such as the Combined Array for Research in Millimeter-wave Astronomy (CARMA; \citealt{woo04,wri10}), the Atacama Large Millimeter/submillimeter Array (ALMA; \citealt{igu09}), and the Very Long Baseline Array (VLBA). The following analysis applies to both of the aforementioned scenarios although for purposes of clear demonstration, we will choose an example with large differences across antennas, making it closer to the second scenario.

We look at the contribution to the image from the visibility, $V_{ab}$, from each antenna pair. While imaging, EPIC introduces a weighting to each antenna during gridding as given by equations~\ref{eqn:dirty-image-FX} and \ref{eqn:e-field-conv}, and equivalently, the weighted visibility from an antenna pair $ab$ projected on the grid is:
\begin{align}
  V^\prime_{ab}(\mathbf{r},f) &= B^{\textrm{G}\star}_{ab}(\mathbf{r}-\mathbf{r}_{ab},f)\,V_{ab},
\end{align}
where, $B^\textrm{G}_{ab}(\mathbf{r}-\mathbf{r}_{ab},f)$ is obtained by the spatial cross-correlation of weighting kernels associated with the individual antennas. The superscript `$\textrm{G}$' denotes the weighting introduced in analysis during the gridding process. We will use superscript `I' to denote the inherent weighting applied by the instrument during the measurement process in obtaining $V_{ab}$ from the true sky model. Optimal imaging requires $W^\textrm{G}_a = W^\textrm{I}_a$ \citep{mor09,mor11}. However, we keep the two superscripts separate here to describe output images when the array is wrongly assumed to be homogeneous.

The sky response of the weighted visibility, referred to as {\it fringes}, is:
\begin{align}
  \mathcal{I}^\prime_{ab}(\hat{\mathbf{s}},f) &= \mathcal{W}^{\textrm{G}\star}_a(\hat{\mathbf{s}},f)\,\mathcal{W}^\textrm{G}_b(\hat{\mathbf{s}},f)\,V_{ab}\,e^{i 2\pi f\mathbf{r}_{ab}\!\cdot\,\hat{\mathbf{s}}/c}.
\end{align}
Since all quantities in the celestial plane (calligraphic fonts) are functions of position, $\hat{\mathbf{s}}$, and frequency, $f$, we drop writing this dependence explicitly hereafter.

The image output of EPIC is the average of these weighted fringes from all antenna pairs:
\begin{align}\label{eqn:wt-dirty-image-EPIC}
  \mathcal{I}^\prime &= \frac{1}{\Nant(\Nant-1)}\,\sum_{\substack{ab\\a\ne b}}\,\mathcal{I}^\prime_{ab}.
\end{align}

We define the effective attenuation of the dirty image due to instrumental and gridding weights as:
\begin{align}\label{eqn:effective-weighting}
  \mathcal{W}_\textrm{eff} &= \mathcal{I}^\prime\, / \,\mathcal{I}_\textrm{D}^\textrm{iso},
\end{align}
where, $\mathcal{I}_\textrm{D}^\textrm{iso}$ is the dirty image with no beam weighting (isotropic, uniform weighting of the sky).

For a homogeneous array, $W^\textrm{G} \equiv W^\textrm{G}_a \equiv W^\textrm{G}_b$. Thus,  
\begin{align}\label{eqn:wt-dirty-image-homogeneous}
  \mathcal{I}^\prime &= \left|\mathcal{W}^\textrm{G}\right|^2\,\left|\mathcal{W}^\textrm{I}\right|^2\,\mathcal{I}_\textrm{D}^\textrm{iso}.
\end{align}
Consistent with \citet{mor09}, the dirty image, which is already attenuated by the instrumental power pattern inherent in the measurement process, gets further attenuated by the power pattern introduced in the gridding step in EPIC imager.

For a heterogeneous array, 
\begin{align}\label{eqn:wt-dirty-image-EPIC-decomp}
  \mathcal{I}^\prime &= \frac{1}{\Nant(\Nant-1)}\,\sum_{pq}\sum_{\substack{ab\\a\ne b\\a\in p, b\in q}}\,\mathcal{I}^\prime_{ab},
\end{align}
where, $ab$ indexes pairs of antennas $a$ and $b$, $pq$ indexes the different types of antenna pairs, with $p$ and $q$ indexing antenna types (note that $ab$ and $pq$ are simply indices in our notation and are not to be interpreted as products of the individual indices). Thus, equation~\ref{eqn:wt-dirty-image-homogeneous} is re-written as the sum of fringes over all unique antenna pairs in an antenna pair type (inner sum) and subsequently summed over all antenna pair types (outer sum). 

For purposes of demonstration in this study, we assume that the spatial distribution of antenna pairs in each antenna pair type is similar and thus results in a similar dirty image, $\mathcal{I}_\textrm{D}^\textrm{iso}$. This usually holds when antennas under each antenna type are chosen randomly or identically such that differences in the synthesized beams are insignificant. Then, equation~\ref{eqn:wt-dirty-image-EPIC-decomp} can be approximated as:
\begin{align}\label{eqn:wt-dirty-image-EPIC-decomp-approx1}
  \mathcal{I}^\prime &\approx \frac{\mathcal{I}_\textrm{D}^\textrm{iso}}{\Nant(\Nant-1)}\,\sum_{pq}\sum_{\substack{ab\\a\ne b\\a\in p, b\in q}} \mathcal{W}^{\textrm{G}\star}_a\,\mathcal{W}^\textrm{G}_b\,\mathcal{W}^\textrm{I}_a\,\mathcal{W}^{\textrm{I}\star}_b,
\end{align}
where, $\mathcal{I}_\textrm{D}^\textrm{iso}$ is the same for all terms and has been pulled out of the summations. For all antennas indexed by $a$ that belong to a particular type $p$, we replace $\mathcal{W}_a$ with $\mathcal{W}_{(p)}$ and is applicable to those arising from both instrumental (superscript `$I$') and gridding (superscript `$G$') origins. If $n_{pq}$ is the number of antenna pairs in the antenna pair type $pq$ (unrelated to the noise quantity also denoted by $n$ in \S\ref{sec:math}), $\mathcal{W}_\textrm{eff}$ can be simplified as:
\begin{align}\label{eqn:wt-dirty-image-EPIC-decomp-approx2}
  \mathcal{W}_\textrm{eff} &\approx \frac{1}{\Nant(\Nant-1)}\,\sum_{pq}n_{pq}\,\mathcal{W}^{\textrm{G}\star}_{(p)}\,\mathcal{W}^\textrm{G}_{(q)}\,\mathcal{W}^\textrm{I}_{(p)}\,\mathcal{W}^{\textrm{I}\star}_{(q)}.
\end{align}
In an optimally weighted image, $\mathcal{W}^\textrm{G}_{(p)} \equiv \mathcal{W}^\textrm{I}_{(q)}$. Thus,
\begin{align}\label{eqn:effective-weighting-optimal}
  \mathcal{W}_\textrm{eff}^\textrm{opt} &\approx \frac{1}{\Nant(\Nant-1)}\,\sum_{pq} n_{pq}\,\left|\mathcal{W}^\textrm{I}_{(p)}\right|^2\left|\mathcal{W}^\textrm{I}_{(q)}\right|^2,
\end{align}
where, superscript `$\textrm{opt}$' denotes `optimal'.

For purposes of illustration, we consider a simple scenario in which there are two antenna types each containing roughly equal numbers of antennas with similar layouts but the analysis is performed under the wrong assumption that all antennas are identical and are all of the first type. In an optimally weighted image, the optimal effective weighting given by equation~\ref{eqn:effective-weighting-optimal} will be:
\begin{align}\label{eqn:effective-weighting-optimal-example}
  \mathcal{W}_\textrm{eff}^\textrm{opt} &\approx \frac{1}{\Nant(\Nant-1)}\,\left[n_1(n_1-1)\,\left|\mathcal{W}^\textrm{I}_{(1)}\right|^4 \right.\nonumber\\
  &\qquad\qquad\qquad\qquad + n_2(n_2-1)\,\left|\mathcal{W}^\textrm{I}_{(2)}\right|^4 \nonumber\\
  &\qquad\qquad\qquad\qquad \left. + n_1n_2\,\left|\mathcal{W}^\textrm{I}_{(1)}\right|^2\,\left|\mathcal{W}^\textrm{I}_{(2)}\right|^2\right].
\end{align}
On the other hand, $\mathcal{W}^\textrm{G}_{(1)} = \mathcal{W}^\textrm{I}_{(1)}$ and $\mathcal{W}^\textrm{G}_{(2)} = \mathcal{W}^\textrm{I}_{(1)}$ for an image obtained in the erroneous case in which all antennas are assumed to be identical. It will be sub-optimal. From equation~\ref{eqn:wt-dirty-image-EPIC-decomp-approx2}, the effective weighting will be:
\begin{align}\label{eqn:effective-weighting-suboptimal-example}
  \mathcal{W}_\textrm{eff}^\textrm{sub} &\approx \frac{1}{\Nant(\Nant-1)}\,\left[n_1(n_1-1)\,\left|\mathcal{W}^\textrm{I}_{(1)}\right|^4\right.\nonumber\\
  &\qquad\qquad\qquad\qquad + n_2(n_2-1)\,\left|\mathcal{W}^\textrm{I}_{(1)}\right|^2\,\left|\mathcal{W}^\textrm{I}_{(2)}\right|^2 \nonumber\\
  &\qquad\qquad\qquad\qquad \left. + n_1n_2\,\left|\mathcal{W}^\textrm{I}_{(1)}\right|^2 \,\mathcal{W}^\textrm{I}_{(1)}\,\mathcal{W}^{\textrm{I}\star}_{(2)}\right].
\end{align}
Finally, since no {\it a priori} information will be available about the heterogeneous array in case of the erroneous assumption, the effective weighting in the image will be believed to be $\mathcal{W}_\textrm{eff}^\textrm{err} = \left|\mathcal{W}_{(1)}^\textrm{I}\right|^4$. 

Estimates of the dirty images with uniform, isotropic weighting in the optimal and erroneous cases will be obtained by dividing the respective weighted images by their assumed effective weights $\mathcal{W}_\textrm{eff}^\textrm{opt}$ and $\mathcal{W}_\textrm{eff}^\textrm{err}$ respectively as:
\begin{align}
  \widehat{\mathcal{I}}_\textrm{D}^\textrm{iso} &\approx \begin{cases}
    \mathcal{I}_\textrm{D}^\textrm{iso} &\textrm{, \qquad optimal case} \\
    \mathcal{I}_\textrm{D}^\textrm{iso}\,\frac{\mathcal{W}_\textrm{eff}^\textrm{sub}}{\mathcal{W}_\textrm{eff}^\textrm{err}} &\textrm{, \qquad erroneous case}
  \end{cases}
\end{align}

If $\mathcal{I}_\textrm{M}$ is the sky model, then $\mathcal{I}_\textrm{D}^\textrm{iso}$ can be expressed as:
\begin{align}\label{eqn:img-uncertainties}
  \mathcal{I}_\textrm{D}^\textrm{iso} &= \mathcal{I}_\textrm{M} + \delta\mathcal{I}_\textrm{S} + \delta\mathcal{I}_\textrm{M},
\end{align}
where the second and third terms on the right hand side denote fluctuations due to sidelobes and those intrinsic to the model, respectively. We define the normalized estimates of the sky model $\widehat{\mathcal{I}}_\textrm{M}^\textrm{norm} \equiv \widehat{\mathcal{I}}_\textrm{D}^\textrm{iso} / \mathcal{I}_\textrm{M}$. Ideally, $\widehat{\mathcal{I}}_\textrm{M}^\textrm{norm}$ should be consistent with unity and not exceed beyond levels of uncertainty as determined by equation~\ref{eqn:img-uncertainties}.

\begin{figure*}
\subfloat[][Optimally weighted EPIC image]{\label{fig:nonphysical-image}\includegraphics[width=0.5\linewidth]{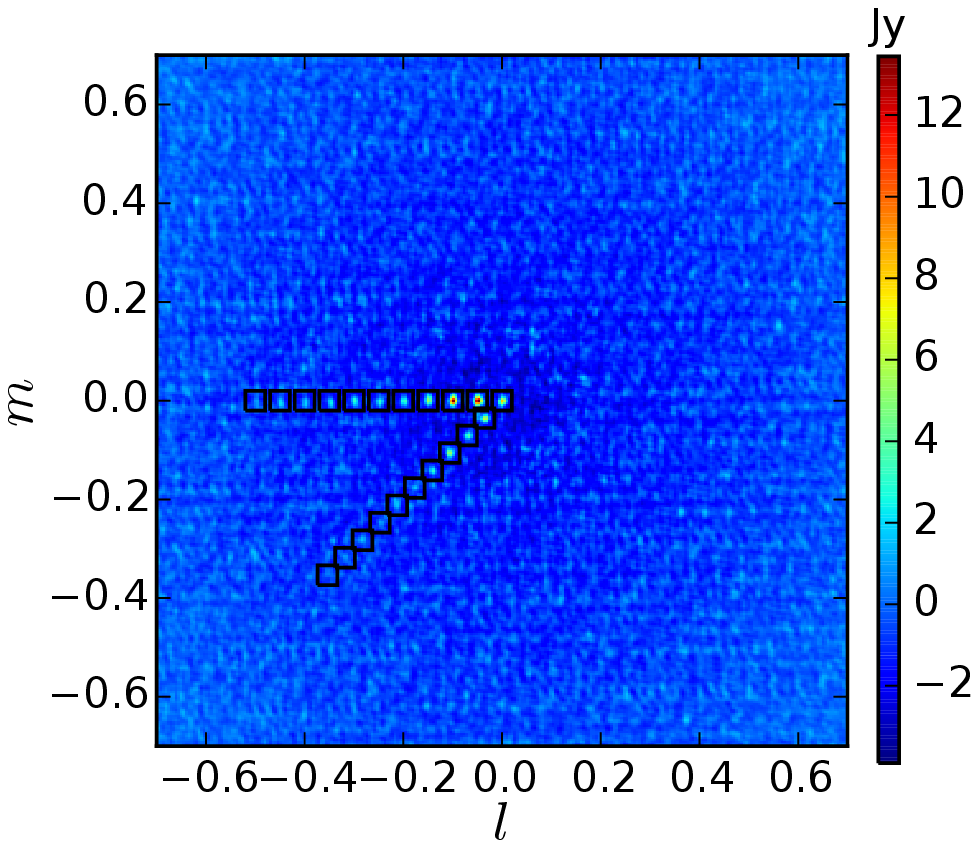}}
\subfloat[][Estimates of normalized radial flux density profile]{\label{fig:norm-flux-density}\includegraphics[width=0.42\linewidth]{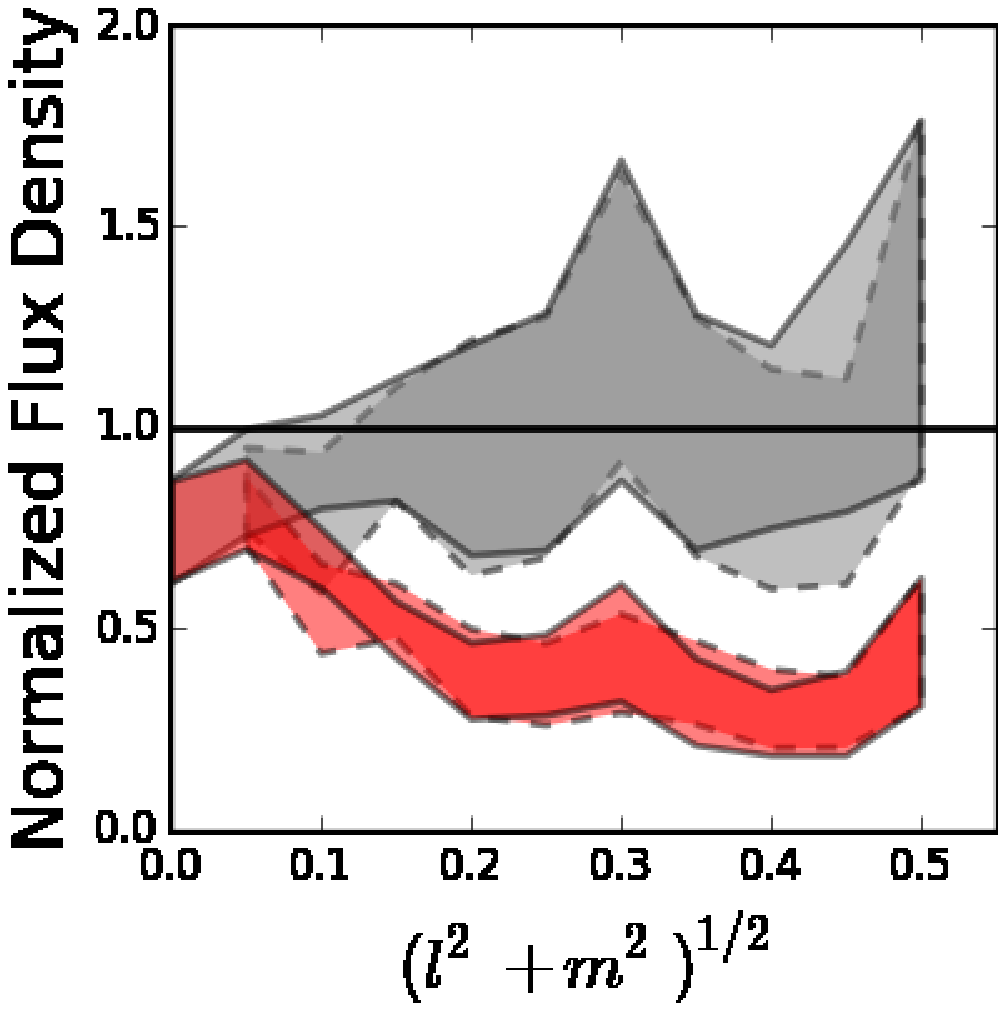}}
\caption{(a)~Optimally weighted image, $\mathcal{I}^\prime$, of simulated sky model obtained by EPIC using MOFF algorithm after averaging over entire bandwidth and duration of simulated observation. 51 antennas in the inner core of the MWA layout were used with uniformly illuminated square apertures. Roughly half of them were squares with 1.1~m sides (first type) and the rest with 6.6~m (second type). Point sources were simulated along westward and south-westward directions with random flux densities brighter than 10~Jy. Black boxes shown around point sources are for purposes of collecting image statistics. (b)~Normalized flux density estimates, $\widehat{\mathcal{I}}_\textrm{M}^\textrm{norm}$, as a function of distance from image centre. Bands indicate an approximate ``$1\sigma$'' uncertainty arising from sidelobe confusion and intrinsic randomness of electric fields simulated for point sources in the sky model. Grey and red bands correspond to optimal and erroneous cases respectively, where in the latter, all antennas were erroneously assumed to be identical and of the first type. The dashed and solid line styles correspond to point sources in the westward and south-westward directions respectively. In the optimal case, the normalized estimates are consistent with unity to within ``$1\sigma$'' in almost all regions of the image, whereas in the erroneous case, they are mis-estimated (underestimated) with high significance ($\gtrsim 5\sigma$) in most regions of the image. There is no significant difference in estimates between the two different directions of point sources studied.}
\label{fig:versatility}
\end{figure*}

In our simulations, the antenna layout, duration of observation, centre frequency and bandwidth used are the same as in \S\ref{sec:sim}. Antennas have square apertures. Those of the first and second types, which have roughly equal numbers of antennas with similar layouts, have 1.1~m and 6.6~m on their sides respectively. $\mathcal{I}_\textrm{M}$ consists of point sources of random flux densities brighter than 10~Jy along westward and south-westward directions as seen in Fig.~\ref{fig:nonphysical-image}, which shows an optimally weighted image, $\mathcal{I}^\prime$, obtained with EPIC using the MOFF algorithm after averaging over the entire bandwidth and duration of simulated observation. 

$\widehat{\mathcal{I}}_\textrm{D}^\textrm{iso}$ was estimated at the point source locations while the sidelobe levels were estimated from boxes three times bigger on the side than those shown using the `median absolute deviation' metric which is resistant to outliers (other point sources in the sky model in our case). The intrinsic fluctuations in the modeled point sources were obtained statistically over time. The net uncertainty was determined by adding the levels of sidelobe fluctuations and intrinsic source fluctuations in quadrature. The normalized estimates of the sky model, $\widehat{\mathcal{I}}_\textrm{M}^\textrm{norm}$, and the corresponding ``$1\sigma$'' error bars were obtained by dividing $\widehat{\mathcal{I}}_\textrm{D}^\textrm{iso}$ and the net uncertainty by the respective effective weights in either case. 

Fig.~\ref{fig:norm-flux-density} shows $\widehat{\mathcal{I}}_\textrm{M}^\textrm{norm}$ including the uncertainty as a function of distance from the centre of the image. The grey and red bands denote estimates from the optimal and the erroneous cases, respectively. The solid and dashed lines correspond to point sources in the south-westward and westward directions respectively. It can be clearly seen that the estimates in the optimal case in almost all parts of the image are consistent with unity to within $1\sigma$. In complete contrast, in the erroneous case, the estimated values appear to be underestimated and deviate from the ideal value of unity with very high significance ($\gtrsim 5\sigma$) in most parts of the image. The reason for underestimation is because nearly half the antennas are assumed to be smaller than they actually are and thus results in dividing by overestimated $\mathcal{W}_\textrm{eff}^\textrm{err}$. These trends do not seem to be affected by the placement of sources along westward and south-westward directions. 

This demonstrates that images from EPIC can robustly image data from heterogeneous arrays while being mathematically optimal. Besides being sub-optimal, erroneous assumptions about the homogeneity of an array can lead to significant and systematic mis-estimates (an underestimate in the above example) of the sky model.

\section{Analysis and Feasibility}\label{sec:analysis}

We now investigate the feasibility of implementing the EPIC imager on current and future radio telescopes. 

\subsection{Processing Volumes}

We have profiled the core routines of EPIC line-by-line for various ranges of parameters such as antenna filling fraction, maximum baseline length, bandwidth and frequency resolution, integration time-scale, etc. for HERA antenna layouts which are highly compact. However, we note that in general, the hardware and optimization of routines in place will determine the relative speeds of the different stages in the pipeline. 

Of all the steps in the MOFF pipeline that are repeated for every writeout from the F-engine, the most expensive step even for dense HERA layouts is found to be the spatial two-dimensional FFT in the imaging stage relative to applying the sparse matrix gridding convolution, squaring or time-averaging. For instance, even in the conservative dense array layout scenario that makes these other stages even more expensive, the gridding convolution, squaring and time-averaging take up only $\lesssim 20$\%, $\lesssim 20$\% and $\lesssim 5$\% respectively of the total processing time while the spatial Fourier transform takes up $\gtrsim 55$\% of the total time. With sparser arrays the gridding process will be be even faster. 

In visibility-based imaging, the predominant computational cost is at the X-engine requiring $\Nant(\Nant-1)/2$ complex multiplications per channel at every A/D writeout time-scale. 

In the following discussions, we will assume that the computational cost for the MOFF imaging is determined by the spatial Fourier transform while that for visibility-based imaging comes from the cross-correlations. However, if non-linearities such as non-coplanarity of baselines \citep{cor08} and wide-field phenomena like the {\it pitchfork} effect \citep{thy15a,thy15b} are to be corrected for, the antenna illumination footprint will become less compact in the measurement plane and can result in a costlier gridding process.

The number of complex multiplications and accumulations in the spatial Fourier transform implemented via FFT is $\approx \beta\,(4\Ngrid)\log_2(4\Ngrid)$ where $\Ngrid$ is the number of pixels on the grid, the factor 4 accounts for the increase in the number of pixels as a result of zero-padding before the spatial Fourier transform, and $\beta$ is a constant that depends on the implementation of twiddle FFT algorithms \citep{bri74}. In our study, we set $\beta=5$, a value\footnote{http://www.fftw.org/speed/method.html} much more conservative than was indicated in \citet{mor11}. We set the number of complex multiplications in the X-engine in visibility-based imaging to $\Nant(\Nant-1)/2$.

We consider a variety of current and planned radio telescopes. Their antenna layouts are summarized in Table~\ref{tab:antenna-layouts}. The size of the layout gives the maximum baseline $b_\textrm{max}$. The grid spacing is determined by the science goals of the experiment in general. For our purpose, we assume a typical requirement that only the field of view of the antenna is to be imaged. This sets the grid spacing to be equal to the size of the antenna, $A_\textrm{a}$. Hence, $\Ngrid\simeq b_\textrm{max}^2/A_\textrm{a}$. 

\begin{table}
  \scriptsize
  \centering
  \caption{Radio telescopes and array layouts.}
  \label{tab:antenna-layouts}
  \begin{threeparttable}
  \begin{tabular}{ccccc} 
    \hline
    Telescope & Core & Number of & Antenna & Frequency \\
              & size & Antennas & size & \\
              & $b_\textrm{max}$ (m) & $\Nant$ & $A_\textrm{a}$ (m$^2$) & $f_0$ (MHz) \\
    \hline
    MWA-112\tnote{a} & 1400 & 112 & 16 & 150 \\
    MWA-240\tnote{a} & 1400 & 240 & 16 & 150 \\
    MWA-496\tnote{a} & 1400 & 496 & 16 & 150 \\
    MWA-112\tnote{a} & 1400 & 1008 & 16 & 150 \\
    LOFAR-LC\tnote{b} & 3500 & 24 & 5809 & 50 \\
    LOFAR-HC\tnote{b} & 3500 & 48 & 745 & 150 \\
    LWA1 & 100 & 256 & 10 & 50 \\
    LWA-OV\tnote{c} & 200 & 256 & 10 & 50 \\
    HERA-19 & 70 & 19 & 154 & 150 \\
    HERA-37 & 98 & 37 & 154 & 150 \\
    HERA-331 & 294 & 331 & 154 & 150 \\
    HERA-6769\tnote{d} & 1330 & 6769 & 154 & 150 \\
    SKA1-LC\tnote{e} & 1000 & 750 & 962 & 150 \\
    SKA1-LCD\tnote{f} & 1000 & 192,000 & 2 & 150 \\
    CHIME & 100 & 1280 & 8 & 600 \\
    HIRAX\tnote{g} & 200 & 1024 & 28 & 600 \\
    \hline
  \end{tabular}
  \begin{tablenotes}
    \item[a] MWA-N denotes N tiles in the specified core diameter
    \item[b] LC and HC denotes low band and high band stations inside the specified core diameter 
    \item[c] Owens Valley LWA
    \item[d] Hypothetically chosen to have a total collecting area of 1~km$^2$
    \item[e] This is the number of beamformed stations expected to be in the core, roughly three-fourths of the total number
    \item[f] All dipoles inside the core are used as independent elements without station beamforming
    \item[g] Hydrogen Intensity mapping and Real-time Analysis eXperiment
  \end{tablenotes}
  \end{threeparttable}
\end{table}

\begin{figure}
  \includegraphics[width=\columnwidth]{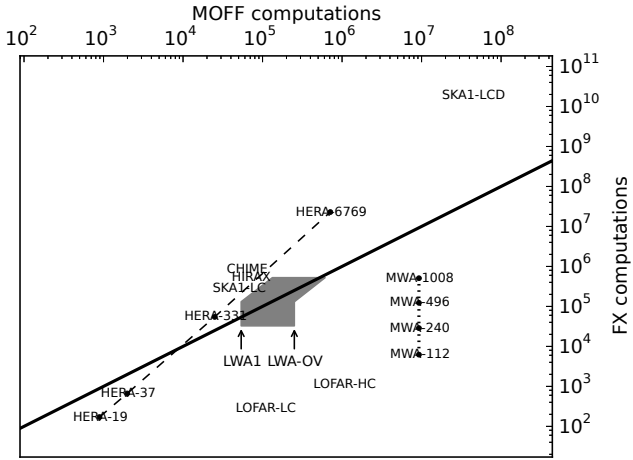}
  \caption{Current and planned instruments in parameter space of number of complex multiplications and additions. The solid line is the boundary at which the number of operations with MOFF and visibility-based imaging are equal. MOFF imaging is more efficient for telescopes occupying the left of this line and vice versa. CHIME, HIRAX, SKA1-LC, SKA1-LCD and all the HERA layouts except HERA-19 and HERA-37 lie in the parameter space favoured by MOFF imaging. The dashed line shows the projected trajectory of hypothetical expanded HERA layouts. The dotted line similarly shows hypothetical expanded MWA layouts with more tiles added in the same core. The grey shaded area denotes the projected trajectory of the LWA bounded by LWA1 (left edge), LWA-OV (right edge), current layout (bottom) and a four-fold increase in the number of elements within a 50\% increase in the core size (top). Current instruments such as MWA and LOFAR fall in a region favoured by visibility-based imaging due to sparse layouts.}
  \label{fig:parameter-space-computations-instruments}
\end{figure}

Fig.~\ref{fig:parameter-space-computations-instruments} shows the number of complex operations per frequency channel per cross-polarization per integration time-scale. Telescopes that fall to the left of the solid line indicate MOFF imaging is computationally more efficient than visibility-based imaging. All HERA layouts except HERA-19 and HERA-37 are in a region of parameter space where MOFF imaging holds the advantage. The dashed line showing future trajectory of HERA-like systems will be clearly favoured by MOFF imaging. The dotted line is similarly a hypothetical trajectory for the MWA with more tiles added inside the same core diameter. The grey shaded area is for a projected LWA expansion and is also predominantly in the region favouring MOFF imaging. It is bounded by the LWA1 and LWA-OV on the left and right respectively. The current (see Table~\ref{tab:antenna-layouts}) and a hypothetical expanded layout with a four-fold increase in number of elements over a 50\% increase in $b_\textrm{max}$ provide the bounds at the bottom and top respectively. Current instruments such as MWA and LOFAR lie in parameter space favouring visibility-based imaging due to the sparseness of their layouts. 

\begin{figure}
  \includegraphics[width=\columnwidth]{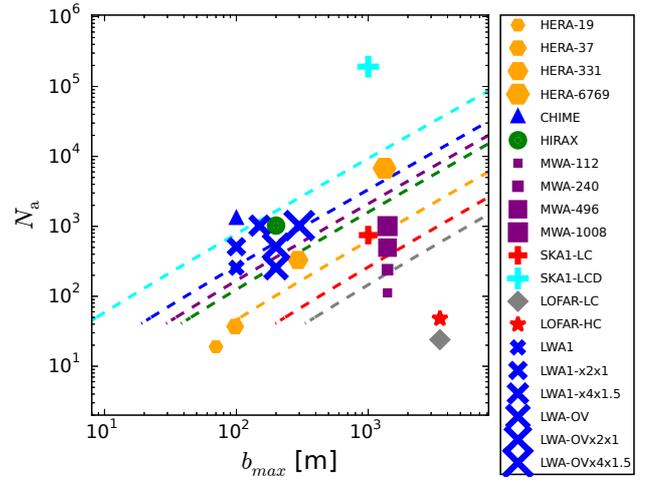}
  \caption{Current and planned instruments in parameter space of baseline length and number of antennas. Lines of different colours denote different classes of antenna sizes (cyan - 1~m$^2$, blue - 7~m$^2$, purple - 16~m$^2$, green - 28~m$^2$, orange - 150~m$^2$, red - 740~m$^2$, grey - 5900~m$^2$). Lines of each colour denote the boundary to the left of which MOFF algorithm is favoured for the corresponding antenna size. These lines shift rightward with increasing antenna size. The different antennas are colour coded by roughly the class of antenna size they fall into. Thus symbols of one colour falling to the left of a line of the same colour indicate MOFF imaging is advantageous for those telescopes and vice versa. For example, MOFF imaging is favoured in HERA-331 and HERA-6769 because they lie to the left of the orange line but not so in cases of HERA-19 and HERA-37.}
  \label{fig:parameter-space-bll-nant-instruments}
\end{figure}

We now consider antenna array layouts described by three essential quantities in radio interferometry, namely, maximum baseline length, number of antennas, and the size of each antenna. Fig.~\ref{fig:parameter-space-bll-nant-instruments} shows the boundaries where the ratio of the number of computations required with visibility-based imaging relative to MOFF imaging is unity. The different coloured lines correspond to different antenna sizes (cyan - 1~m$^2$, blue - 7~m$^2$, purple - 16~m$^2$, green - 28~m$^2$, orange - 150~m$^2$, red - 740~m$^2$, grey - 5900~m$^2$). Dashed lines of each colour denote the boundary to the left of which the MOFF algorithm is favoured for the corresponding antenna size and vice versa for visibility-based imaging.

There is an upper limit set by the maximum number of antennas for each antenna size that can be densely packed inside various baseline lengths but is not shown on this figure to avoid crowding. We note that as antenna size increases the maximum number of antennas for a dense packing as a function of baseline length decreases. Hence, this upper limit shifts rightward as antenna size increases. Similarly, with increase in antenna size, $\Ngrid$ also decreases when field of view imaging is achieved with an increased grid spacing equal to antenna size and hence lowers the amount of computations required with the MOFF algorithm. This shifts the dashed lines rightward as well.

The different antennas are colour coded by roughly the class of antenna size they fall into. Thus symbols of one colour that lie to the left of same coloured line indicate MOFF imaging is favoured for those telescopes and vice versa. For example, MOFF imaging is favoured in HERA-331 and HERA-6769 because they lie to the left of the orange line but not so in cases of HERA-19 and HERA-37. A majority of the next-generation radio telescopes, namely, HERA-331 and its future expanded versions, SKA1-LC, SKA1-LCD, HIRAX, and CHIME will fall in the regime where MOFF imaging will be desirable. LWA1 and LWA-OV are already very close to the dividing line. Their hypothetical expansions,\footnote{LWA1-x2x1 and LWA1-x4x1.5, and LWA-OVx2x1 and LWA-OVx4x1.5 denote two-fold and four-fold increase in number of antennas within a core diameter that is 1 and 1.5 times the current size of 100~m and 200~m respectively for LWA1 and LWA-OV.} will be in the computational regime favoured by MOFF. For a fixed baseline length, regions favouring the MOFF algorithm tend to be towards large $\Nant$ indicating large-N dense array layouts with smaller antenna elements are best suited for deploying EPIC.

\subsection{Data Throughput}

We elaborate on the I/O data rates required with the MOFF and visibility-based algorithms. This is particularly relevant in the context of radio transient detection. 

Implementation of the MOFF algorithm with EPIC yields calibrated images on time-scales of the output generated by the digitizer and is set by the inverse of the frequency channel width. These calibrated images are accumulated and averaged to a certain time-scale depending on science or hardware requirements, or when the sky has rotated significantly, whichever is lesser. In visibility-based algorithms, the visibilities are accumulated and averaged to this time-scale before images are produced. Thus the data throughput (in samples per second) per cross-polarization with MOFF and X-engine outputs are: 
\begin{align}
  r_\textrm{MOFF} &\sim \frac{4\Ngrid}{\Delta t} \left(\frac{\Delta B}{\Delta f}\right) \\
  r_\textrm{X} &\sim 2\,\frac{\Nant(\Nant-1)/2}{\Delta t} \left(\frac{\Delta B}{\Delta f}\right),
\end{align}
where, the factor 4 in the expression for $r_\textrm{MOFF}$ accounts for imaging after zero-padding the gridded electric fields, the leading factor of 2 in the expression for $r_\textrm{X}$ accounts for the real and imaginary parts of the complex visibilities, $\Delta B$ is the bandwidth, $\Delta f$ is the frequency resolution, and $\Delta t$ is the time-scale over which the transient phenomenon is sampled and the data (images or visibilities) are averaged to.

Though a full understanding of the FRB phenomena is yet to emerge, there are indications the time-scales of FRB objects are $\Delta t \sim 1$--10~ms \citep{tho13}. For a telescope like HERA, $\Delta B \simeq 100$~MHz, $\Delta f \simeq 100$~kHz. For HERA-331, $\Nant=331$ and with a grid spacing to image the field of view, $\Ngrid \simeq 441$ or $\Ngrid \simeq 1024$ if $\Ngrid$ is preferred as a power of 2 in either direction. Using 8 bytes for each floating point sample in the MOFF images and 4 bytes each for real and imaginary parts of visibility samples, the throughputs are $r_\textrm{MOFF} \lesssim 3$~GB~s$^{-1}$ and $r_\textrm{X} \simeq 41$~GB~s$^{-1}$. For HERA-37, $r_\textrm{MOFF} \lesssim 190$~MB~s$^{-1}$ and $r_\textrm{X} \simeq 0.5$~GB~s$^{-1}$. In such a case, The X-engine throughput corresponds to an extreme rate of $\simeq 1.8$~TB an hour per cross-polarization. Conversely, for the same data throughput, the MOFF algorithm can sample even shorter time-scales. 

Table~\ref{tab:data-rates} shows these data rates for some of the current and planned telescopes for $\Delta t=10$~ms. In almost all cases listed, even with conservative estimates, the MOFF algorithm provides very economic data throughput for a majority of next generation radio telescopes with a dense layout. The most significant advantage is that calibrated images are also available at no extra cost.

\begin{table}
  \centering
  \caption{Data throughput per cross-polarization for various telescopes with MOFF and X-engine outputs on time-scales of $\Delta t=10$~ms assuming $\Delta B=100$~MHz and $\Delta f=100$~kHz.}
  \label{tab:data-rates}
  \begin{threeparttable}
  \begin{tabular}{ccc} 
    \hline
    Telescope\tnote{a} & $r_\textrm{MOFF}$\tnote{b} & $r_\textrm{X}$ \\
              & (GB~s$^{-1}$)\tnote{c} & (GB~s$^{-1}$)\tnote{c} \\
    \hline
    LWA1 & $\simeq 3$ & $\simeq 24.3$ \\
    LWA-OV & $\simeq 12$ & $\simeq 24.3$ \\
    HERA-19 & $\lesssim 0.19$ & $\simeq 0.13$ \\
    HERA-37 & $\lesssim 0.19$ & $\simeq 0.5$ \\
    HERA-331 & $\lesssim 3$ & $\simeq 41$ \\
    CHIME & $\lesssim 6.1$ & $\simeq 610$ \\
    \hline
  \end{tabular}
  \begin{tablenotes}
    \item[a] Antenna layouts are listed in Table~\ref{tab:antenna-layouts}.
    \item[b] $\Ngrid$ is usually greater than true value because of rounding to the next power of 2 in either direction. Thus $r_\textrm{MOFF}$ is usually lesser than the conservative values listed here.
    \item[c] We assume 8 bytes for each real sample from MOFF images, and 4 bytes each for real and imaginary parts of visibility samples.
  \end{tablenotes}
  \end{threeparttable}
\end{table}

\section{Conclusions}\label{sec:conclusions}

As radio astronomy is entering a new era, advances in instrumentation have to be accompanied by equal advances in processing techniques to manage computational resources. Many future radio telescopes such as the SKA, HERA and LWA are headed towards the large-N dense array layout model for which computational cost from traditional FX/XF correlator-based architecture and visibility-based imaging starts rising steeply. We have provided the first software demonstration of a general purpose imaging algorithm using our generic and efficient EPIC software that is designed to bring this cost down from $\mathcal{O}(\Nant^2)$ to $\mathcal{O}(\Ngrid\log_2 \Ngrid)$. Under the class of direct imaging techniques, ours is one of the most generic -- neither does it place any constraint on the array layout to be on a regular grid nor does it require the antenna array to be homogeneous. We have provided a detailed mathematical framework for imaging with heterogeneous arrays, demonstrated the natural ability of EPIC to robustly image data from such arrays, and shown that wrong assumptions about array homogeneity could result in significant and systematic mis-estimation of the sky model.

Our EPIC package, now publicly available, written in object-oriented Python, is highly modularized and parallelizable. It includes an implementation of the MOFF algorithm in addition to visibility-based software holography imaging and a data simulator for sky models. It is designed to provide a development platform to compare different imaging approaches and serve as a stepping stone for real-life GPU/FPGA-based implementation on telescopes. It has been successfully tested on simulated MWA observations as well as real LWA1 observations from both imaging and calibration viewpoints. 

The MOFF algorithm packaged with EPIC is already found to be most suitable for many present and planned radio telescopes such as the LWA, HERA, CHIME, HIRAX and SKA. In general, MOFF is most suited to operate in the region of parameter space characterized by dense packing of a large number of antennas especially when consisting of a large number of small antenna elements. 

It is seen to have significant savings in data throughput relative to a X-engine based pipeline. A unique advantage is the instantaneous availability of calibrated time-domain images at no extra cost. Hence, it is a compelling candidate for time-domain radio astronomy, e.g. search for and monitoring of transients. Potentially, it could allow on-chip processing thus lowering even further the already relatively low I/O bandwidth shown in Table~\ref{tab:data-rates} except when a transient event is detected. Transient detection pipelines at the backend of EPIC can be fine-tuned to target fast transients such as the Fast Radio Bursts \citep[FRB;][]{tho13} on millisecond time-scales at GHz frequencies or slow transients from planetary and exoplanetary origins at frequencies around 100~MHz. 

Thus, EPIC with the MOFF algorithm packaged is uniquely poised to offer a substantial advantage to imaging with large-N dense arrays typical of next-generation radio telescopes as well as push the frontiers of time-domain astronomy to fill gaps in understanding the science behind phenomena responsible for extreme transient events in the Universe.

In the near future, we plan to demonstrate speed and precision by upgrading EPIC to a GPU-based implementation in order to operate on real-time data and develop a transient trigger and monitor backend. In the meanwhile, we plan to demonstrate imaging with non-coplanar arrays and direction-dependent calibration.

\section*{Acknowledgements}

We thank Larry D'Addario, Gregg Hallinan, Joseph Lazio and Harish Vedantham for their valuable inputs, and Greg Taylor for providing us with LWA1 data. This work has been supported by the National Science Foundation through award AST-1206552. Construction of the LWA has been supported by the Office of Naval Research under Contract N00014-07-C-0147. Support for operations and continuing development of the LWA1 is provided by the National Science Foundation under grant AST-1139974 of the University Radio Observatory program.


\bibliographystyle{mnras}

\begin{thebibliography}{}
\makeatletter
\relax
\def\mn@urlcharsother{\let\do\@makeother \do\$\do\&\do\#\do\^\do\_\do\%\do\~}
\def\mn@doi{\begingroup\mn@urlcharsother \@ifnextchar [ {\mn@doi@}
  {\mn@doi@[]}}
\def\mn@doi@[#1]#2{\def\@tempa{#1}\ifx\@tempa\@empty \href
  {http://dx.doi.org/#2} {doi:#2}\else \href {http://dx.doi.org/#2} {#1}\fi
  \endgroup}
\def\mn@eprint#1#2{\mn@eprint@#1:#2::\@nil}
\def\mn@eprint@arXiv#1{\href {http://arxiv.org/abs/#1} {{\tt arXiv:#1}}}
\def\mn@eprint@dblp#1{\href {http://dblp.uni-trier.de/rec/bibtex/#1.xml}
  {dblp:#1}}
\def\mn@eprint@#1:#2:#3:#4\@nil{\def\@tempa {#1}\def\@tempb {#2}\def\@tempc
  {#3}\ifx \@tempc \@empty \let \@tempc \@tempb \let \@tempb \@tempa \fi \ifx
  \@tempb \@empty \def\@tempb {arXiv}\fi \@ifundefined
  {mn@eprint@\@tempb}{\@tempb:\@tempc}{\expandafter \expandafter \csname
  mn@eprint@\@tempb\endcsname \expandafter{\@tempc}}}

\bibitem[\protect\citeauthoryear{{Bandura} et~al.,}{{Bandura}
  et~al.}{2014}]{ban14}
{Bandura} K.,  et~al., 2014, in Society of Photo-Optical Instrumentation
  Engineers (SPIE) Conference Series. p.~22 (\mn@eprint {arXiv} {1406.2288}),
  \mn@doi{10.1117/12.2054950}

\bibitem[\protect\citeauthoryear{{Beardsley} et~al.,}{{Beardsley}
  et~al.}{2012}]{bea12}
{Beardsley} A.~P.,  et~al., 2012, \mn@doi [\mnras]
  {10.1111/j.1365-2966.2012.20878.x}, \href
  {http://adsabs.harvard.edu/abs/2012MNRAS.425.1781B} {425, 1781}

\bibitem[\protect\citeauthoryear{{Beardsley}, {Thyagarajan}, {Bowman}  \&
  {Morales}}{{Beardsley} et~al.}{2016}]{bea16}
{Beardsley} A.~P.,  {Thyagarajan} N.,  {Bowman} J.~D.,   {Morales} M.~F.,
  2016, preprint, \href {http://adsabs.harvard.edu/abs/2016arXiv160302126B} {}
  (\mn@eprint {arXiv} {1603.02126})

\bibitem[\protect\citeauthoryear{{Bennett}, {Anderson}, {McInnes}  \&
  {Whitaker}}{{Bennett} et~al.}{1976}]{ben76}
{Bennett} J.~C.,  {Anderson} A.~P.,  {McInnes} P.~A.,   {Whitaker} A.~J.~T.,
  1976, \mn@doi [IEEE Transactions on Antennas and Propagation]
  {10.1109/TAP.1976.1141354}, \href
  {http://adsabs.harvard.edu/abs/1976ITAP...24..295B} {24, 295}

\bibitem[\protect\citeauthoryear{{Bhatnagar, S.}, {Cornwell, T. J.}, {Golap,
  K.}  \& {Uson, J. M.}}{{Bhatnagar, S.} et~al.}{2008}]{bha08}
{Bhatnagar, S.} {Cornwell, T. J.} {Golap, K.}  {Uson, J. M.} 2008, \mn@doi
  [A\&A] {10.1051/0004-6361:20079284}, 487, 419

\bibitem[\protect\citeauthoryear{{Bowman} et~al.,}{{Bowman}
  et~al.}{2013}]{bow13}
{Bowman} J.~D.,  et~al., 2013, \mn@doi [\pasa] {10.1017/pas.2013.009}, \href
  {http://adsabs.harvard.edu/abs/2013PASA...30...31B} {30, 31}

\bibitem[\protect\citeauthoryear{{Brigham}}{{Brigham}}{1974}]{bri74}
{Brigham} E.~O.,  1974, {The fast Fourier Transform}

\bibitem[\protect\citeauthoryear{{Bunton}}{{Bunton}}{2004}]{bun04}
{Bunton} J.~D.,  2004, \mn@doi [Experimental Astronomy]
  {10.1007/s10686-005-5661-5}, \href
  {http://adsabs.harvard.edu/abs/2004ExA....17..251B} {17, 251}

\bibitem[\protect\citeauthoryear{Cooley \& Tukey}{Cooley \&
  Tukey}{1965}]{coo65}
Cooley J.~W.,  Tukey J.~W.,  1965, Math. Comput., 19, 297

\bibitem[\protect\citeauthoryear{{Cornwell}, {Golap}  \&
  {Bhatnagar}}{{Cornwell} et~al.}{2008}]{cor08}
{Cornwell} T.~J.,  {Golap} K.,   {Bhatnagar} S.,  2008, \mn@doi [IEEE Journal
  of Selected Topics in Signal Processing] {10.1109/JSTSP.2008.2005290}, \href
  {http://adsabs.harvard.edu/abs/2008ISTSP...2..647C} {2, 647}

\bibitem[\protect\citeauthoryear{Daishido et~al.,}{Daishido
  et~al.}{2000}]{dai00}
Daishido T.,  et~al., 2000, \mn@doi [Proc. SPIE] {10.1117/12.390458}, 4015, 73

\bibitem[\protect\citeauthoryear{{DeBoer} et~al.,}{{DeBoer}
  et~al.}{2016}]{deb16}
{DeBoer} D.~R.,  et~al., 2016, preprint, \href
  {http://adsabs.harvard.edu/abs/2016arXiv160607473D} {} (\mn@eprint {arXiv}
  {1606.07473})

\bibitem[\protect\citeauthoryear{{Ellingson} et~al.,}{{Ellingson}
  et~al.}{2013}]{ell13}
{Ellingson} S.~W.,  et~al., 2013, \mn@doi [IEEE Transactions on Antennas and
  Propagation] {10.1109/TAP.2013.2242826}, \href
  {http://adsabs.harvard.edu/abs/2013ITAP...61.2540E} {61, 2540}

\bibitem[\protect\citeauthoryear{Foster, Hickish, Magro, Price  \&
  Zarb~Adami}{Foster et~al.}{2014}]{fos14}
Foster G.,  Hickish J.,  Magro A.,  Price D.,   Zarb~Adami K.,  2014, \mn@doi
  [Monthly Notices of the Royal Astronomical Society] {10.1093/mnras/stu188},
  439, 3180

\bibitem[\protect\citeauthoryear{{Iguchi} et~al.,}{{Iguchi}
  et~al.}{2009}]{igu09}
{Iguchi} S.,  et~al., 2009, \mn@doi [\pasj] {10.1093/pasj/61.1.1}, \href
  {http://adsabs.harvard.edu/abs/2009PASJ...61....1I} {61, 1}

\bibitem[\protect\citeauthoryear{{Lonsdale}, {Doeleman}, {Cappallo}, {Hewitt}
  \& {Whitney}}{{Lonsdale} et~al.}{2000}]{lon00}
{Lonsdale} C.~J.,  {Doeleman} S.~S.,  {Cappallo} R.~J.,  {Hewitt} J.~N.,
  {Whitney} A.~R.,  2000, in {Butcher} H.~R.,  ed.,  Society of Photo-Optical
  Instrumentation Engineers (SPIE) Conference Series Vol. 4015, Radio
  Telescopes. pp 126--134

\bibitem[\protect\citeauthoryear{{Maneewongvatana} \&
  {Mount}}{{Maneewongvatana} \& {Mount}}{1999}]{man99}
{Maneewongvatana} S.,  {Mount} D.~M.,  1999, eprint arXiv:cs/9901013, \href
  {http://adsabs.harvard.edu/abs/1999cs........1013M} {}

\bibitem[\protect\citeauthoryear{{Mellema} et~al.,}{{Mellema}
  et~al.}{2013}]{mel13}
{Mellema} G.,  et~al., 2013, \mn@doi [Experimental Astronomy]
  {10.1007/s10686-013-9334-5}, \href
  {http://adsabs.harvard.edu/abs/2013ExA....36..235M} {36, 235}

\bibitem[\protect\citeauthoryear{{Morales}}{{Morales}}{2011}]{mor11}
{Morales} M.~F.,  2011, \mn@doi [\pasp] {10.1086/663092}, \href
  {http://adsabs.harvard.edu/abs/2011PASP..123.1265M} {123, 1265}

\bibitem[\protect\citeauthoryear{{Morales} \& {Matejek}}{{Morales} \&
  {Matejek}}{2009}]{mor09}
{Morales} M.~F.,  {Matejek} M.,  2009, \mn@doi [\mnras]
  {10.1111/j.1365-2966.2009.15537.x}, \href
  {http://adsabs.harvard.edu/abs/2009MNRAS.400.1814M} {400, 1814}

\bibitem[\protect\citeauthoryear{{Otobe} et~al.,}{{Otobe} et~al.}{1994}]{oto94}
{Otobe} E.,  et~al., 1994, \pasj, \href
  {http://adsabs.harvard.edu/abs/1994PASJ...46..503O} {46, 503}

\bibitem[\protect\citeauthoryear{Parsons et~al.,}{Parsons et~al.}{2010}]{par10}
Parsons A.~R.,  et~al., 2010, The Astronomical Journal, 139, 1468

\bibitem[\protect\citeauthoryear{{Scott} \& {Ryle}}{{Scott} \&
  {Ryle}}{1977}]{sco77}
{Scott} P.~F.,  {Ryle} M.,  1977, \mn@doi [\mnras] {10.1093/mnras/178.4.539},
  \href {http://adsabs.harvard.edu/abs/1977MNRAS.178..539S} {178, 539}

\bibitem[\protect\citeauthoryear{Tegmark}{Tegmark}{1997a}]{teg97b}
Tegmark M.,  1997a, \mn@doi [Phys. Rev. D] {10.1103/PhysRevD.55.5895}, 55, 5895

\bibitem[\protect\citeauthoryear{{Tegmark}}{{Tegmark}}{1997b}]{teg97a}
{Tegmark} M.,  1997b, \mn@doi [\apjl] {10.1086/310631}, \href
  {http://adsabs.harvard.edu/abs/1997ApJ...480L..87T} {480, L87}

\bibitem[\protect\citeauthoryear{{Tegmark} \& {Zaldarriaga}}{{Tegmark} \&
  {Zaldarriaga}}{2009}]{teg09}
{Tegmark} M.,  {Zaldarriaga} M.,  2009, \mn@doi [\prd]
  {10.1103/PhysRevD.79.083530}, \href
  {http://adsabs.harvard.edu/abs/2009PhRvD..79h3530T} {79, 083530}

\bibitem[\protect\citeauthoryear{Tegmark \& Zaldarriaga}{Tegmark \&
  Zaldarriaga}{2010}]{teg10}
Tegmark M.,  Zaldarriaga M.,  2010, \mn@doi [Phys. Rev. D]
  {10.1103/PhysRevD.82.103501}, 82, 103501

\bibitem[\protect\citeauthoryear{{Thompson}, {Moran}  \& {Swenson}}{{Thompson}
  et~al.}{2001}]{tho01}
{Thompson} A.~R.,  {Moran} J.~M.,   {Swenson} Jr. G.~W.,  2001, {Interferometry
  and Synthesis in Radio Astronomy, 2nd Edition}.
Wiley

\bibitem[\protect\citeauthoryear{{Thornton} et~al.,}{{Thornton}
  et~al.}{2013}]{tho13}
{Thornton} D.,  et~al., 2013, \mn@doi [Science] {10.1126/science.1236789},
  \href {http://adsabs.harvard.edu/abs/2013Sci...341...53T} {341, 53}

\bibitem[\protect\citeauthoryear{{Thyagarajan} et~al.,}{{Thyagarajan}
  et~al.}{2015a}]{thy15a}
{Thyagarajan} N.,  et~al., 2015a, \mn@doi [\apj] {10.1088/0004-637X/804/1/14},
  \href {http://adsabs.harvard.edu/abs/2015ApJ...804...14T} {804, 14}

\bibitem[\protect\citeauthoryear{{Thyagarajan} et~al.,}{{Thyagarajan}
  et~al.}{2015b}]{thy15b}
{Thyagarajan} N.,  et~al., 2015b, \mn@doi [\apjl]
  {10.1088/2041-8205/807/2/L28}, \href
  {http://adsabs.harvard.edu/abs/2015ApJ...807L..28T} {807, L28}

\bibitem[\protect\citeauthoryear{{Tingay} et~al.,}{{Tingay}
  et~al.}{2013}]{tin13}
{Tingay} S.~J.,  et~al., 2013, \mn@doi [\pasa] {10.1017/pasa.2012.007}, \href
  {http://adsabs.harvard.edu/abs/2013PASA...30....7T} {30, e007}

\bibitem[\protect\citeauthoryear{Wijnholds \& van~der Veen}{Wijnholds \&
  van~der Veen}{2009}]{wij09}
Wijnholds S.,  van~der Veen A.-J.,  2009, \mn@doi [Signal Processing, IEEE
  Transactions on] {10.1109/TSP.2009.2022894}, 57, 3512

\bibitem[\protect\citeauthoryear{{Woody} et~al.,}{{Woody} et~al.}{2004}]{woo04}
{Woody} D.~P.,  et~al., 2004, in {Bradford} C.~M.,  et~al., eds,  \procspie
  Vol. 5498, Z-Spec: a broadband millimeter-wave grating spectrometer: design,
  construction, and first cryogenic measurements. pp 30--41,
  \mn@doi{10.1117/12.552446}

\bibitem[\protect\citeauthoryear{{Wright}}{{Wright}}{2010}]{wri10}
{Wright} M.~C.~H.,  2010, in Ground-based and Airborne Telescopes III. p.
  77331B, \mn@doi{10.1117/12.856707}

\bibitem[\protect\citeauthoryear{{Zernike}}{{Zernike}}{1938}]{zer38}
{Zernike} F.,  1938, \mn@doi [Physica] {10.1016/S0031-8914(38)80203-2}, \href
  {http://adsabs.harvard.edu/abs/1938Phy.....5..785Z} {5, 785}

\bibitem[\protect\citeauthoryear{{Zheng} et~al.,}{{Zheng} et~al.}{2014}]{zhe14}
{Zheng} H.,  et~al., 2014, \mn@doi [\mnras] {10.1093/mnras/stu1773}, \href
  {http://adsabs.harvard.edu/abs/2014MNRAS.445.1084Z} {445, 1084}

\bibitem[\protect\citeauthoryear{de Vos, Gunst  \& Nijboer}{de~Vos
  et~al.}{2009}]{dev09}
de Vos M.,  Gunst A.,   Nijboer R.,  2009, \mn@doi [Proceedings of the IEEE]
  {10.1109/JPROC.2009.2020509}, 97, 1431

\bibitem[\protect\citeauthoryear{{van Cittert}}{{van Cittert}}{1934}]{van34}
{van Cittert} P.~H.,  1934, \mn@doi [Physica] {10.1016/S0031-8914(34)90026-4},
  \href {http://adsabs.harvard.edu/abs/1934Phy.....1..201V} {1, 201}

\bibitem[\protect\citeauthoryear{{van Haarlem} et~al.,}{{van Haarlem}
  et~al.}{2013}]{van13}
{van Haarlem} M.~P.,  et~al., 2013, \mn@doi [\aap]
  {10.1051/0004-6361/201220873}, \href
  {http://adsabs.harvard.edu/abs/2013A%26A...556A...2V} {556, A2}

\makeatother
\end{thebibliography}
\input{ms.bbl}


\appendix

\section{Software Architecture}\label{sec:software-modules}

\begin{figure*}
  \includegraphics[width=\linewidth]{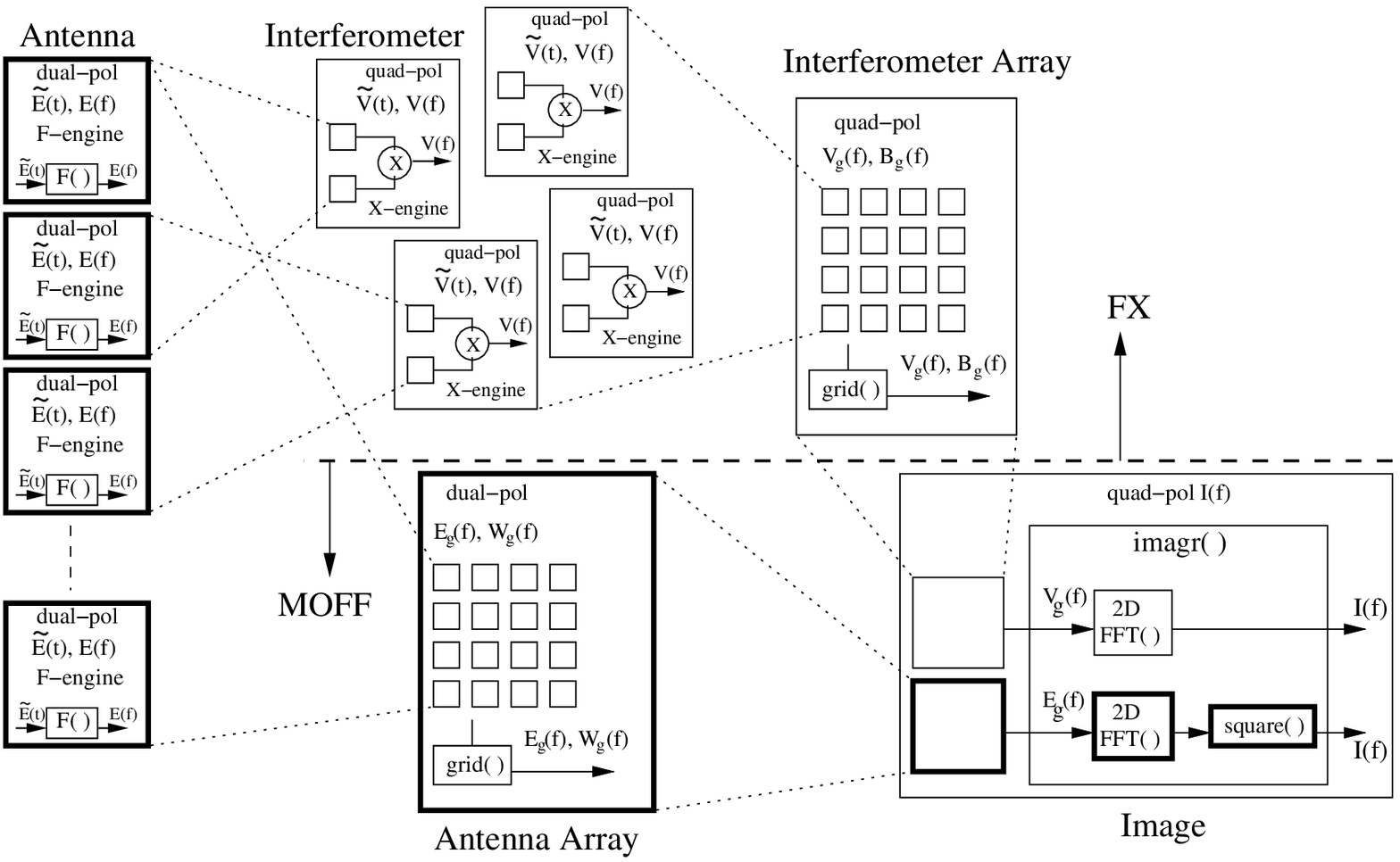}
  \caption{Software architecture of EPIC with core modules, their essential attributes and functions. The antenna module forms the fundamental building block. It consists of electric field time-series and spectra and the F-engine that performs a temporal FFT to obtain electric field spectra from the time-series. The interferometer module is made of a pair of antenna modules. Its main function is the X-engine (FX or XF) to produce visibility spectra. The antenna array module is made of all individual antenna modules as its components and contains collective properties about the antenna subsystems. Its core function is the creation of antenna-to-grid mapping, gridded aperture weights and electric fields. The interferometer array module is very similar in principle to the antenna array module except it operates on cross-correlations and produces gridded visibilities. The image module takes gridded electric fields or visibilities and performs a two-dimensional spatial FFT (and squares the intermediate image in case of the former) to produce output images. Broadly, the MOFF algorithm is implemented by modules below the horizontal dashed line while the visibility-based imaging uses modules above the line. The exact processing pathway implementing the MOFF algorithm is shown in bolded modules.}
  \label{fig:software-modules}
\end{figure*}

EPIC is built using object-oriented programming in Python and is built on carefully crafted modules which closely represent real-life entities in radio interferometer arrays and observations. The essential modules along with their key attributes and methods are illustrated in Fig.~\ref{fig:software-modules}. These modules are described below.

\subsection{Antenna Module}

The antenna module is a fundamental building block upon which all the other modules are built. There is one antenna module per antenna each having attributes -- the propagated electric field time-series, $\widetilde{E}(t)$, and spectrum $E(f)$ for both polarizations. The most important function inside this module is the F-engine that Fourier-transforms time-series electric field data into spectra. 

The other function (not shown in the figure) is to update the data as new data streams in. This can also be parallelized. Another important attribute consists of antenna flags (not shown in the figure) for each polarization appropriate for the data stream being held by the module. 

\subsection{Interferometer Module}

The interferometer module holds the attributes and functions pertaining to a pair of antennas and represents the cross-correlation information obtained from the pair. Its primary attributes are the two antenna modules. It also contains four cross-polarized visibility time-series (even for the FX correlator for diagnostic purposes) and spectra. 

The critical component of the interferometer module is the X-engine. This is essentially a software analog of hardware correlators of real telescope systems. The X-engine can be toggled between two states of operation, namely, the FX and XF modes. The FX mode obtains the electric field spectra, $E(f)$ from the individual antenna modules inside this module and multiplies the two to obtain visibility spectra, $V(f)$. On the other hand, the XF mode cross-correlates the electric field time-series from its Antenna modules to obtain the visibilities as a function of lags, $V_t(\tau)$, which is then Fourier-transformed to obtain $V(f)$. Both modules can operate on dual-polarizations to obtain all four cross-polarizations.

The other attributes (not shown in the figure) are the flags applicable for each cross-polarization for the current data stream. Similar to the antenna module, it has an update function that can update the visibilities $V_t(\tau)$ or $V(f)$ directly rather than through the electric fields of its component antennas. This functionality is to allow EPIC to operate while attached to the backend of traditional correlator systems. This feature is not utilized for purposes of this paper.

This module forms the fundamental unit for the interferometer array module (to be discussed below) and in general for visibility-based correlator and imaging systems. 

\subsection{Antenna Array Module}

The antenna array module consists of all the antenna modules as its attributes and represents the collective properties of its component antennas. By virtue of holding each antenna data independently in their respective modules, the F-engine for the entire array can be distributed to the F-engines of the component antenna modules thus achieving a highly parallelized F-engine while emulating real telescope systems.

The primary attributes held by this module are the antenna aperture illumination weights and electric fields projected on the grid using the gridding convolution method described above and implemented by the gridding function in this module. Significant parts of the antenna-to-grid mapping and gridding convolution are parallelizable across antennas and frequencies.

Individual antenna flags are carried over as additional weights to be applied to the gridded aperture illumination and electric fields. A series of data streams can be stacked up to take advantage of the array optimization available in Python. This module is also equipped to manage dual-polarization. 

\subsection{Interferometer Array Module}

Similar to the antenna array module, the interferometer array module consists of individual interferometer modules. It can parallelize the correlator operations by distributing the X-operation over the X-engines of its component interferometer modules. The interferometer-to-grid mapping and gridding convolution are very similar in nature to that of the antenna array module. Flag-based grid weights, stacking and ability to handle all four cross-polarizations are built into this module. 

\subsection{Image Module}

The image module is built as a general purpose module that can switch between operating on gridded electric fields or visibilities. At its heart, it consists of a two-dimensional spatial FFT where the padding can be specified by the user to control the resolution in the output images. In case of MOFF imaging, there is an additional step of squaring the holographic electric field images. 

Besides its core functions of spatial Fourier transform and squaring, it can stack, accumulate and average images, and optionally remove the antenna auto-correlations centred around the zero-spacing pixel in the $uv$ plane. It also handles all four cross-polarization products. Currently, it supports writing data out in standard FITS format. 

\section{Mathematical Framework for Imaging with Heterogeneous Arrays}\label{sec:math-versatility}

We present a methodology to understand the effective angular weighting in the image obtained with EPIC from a heterogeneous array and compare it with that from existing imaging applications which assume a homogeneous array (all antennas are identical). 

Dropping the noise term and time dependence, the visibility measured by an antenna pair can be written from equation~\ref{eqn:measurement-eqn-1} as:
\begin{align}\label{eqn:apndx-measurement-eqn-2}
  V_{ab}(f) &= \int \mathcal{W}^\textrm{I}_a(\hat{\mathbf{s}},f)\mathcal{W}^{\textrm{I}\star}_b(\hat{\mathbf{s}},f)\,\mathcal{I}(\hat{\mathbf{s}},f)\,e^{-i 2\pi f\mathbf{r}_{ab}\!\cdot\,\hat{\mathbf{s}}/c}\,\dif\Omega,
\end{align}
where, superscript `$\textrm{I}$' in the $\mathcal{W}$ term indicates the weighting is inherently introduced by the instrument during the measurement process. The beam-weighted dirty image is:
\begin{align}
  \mathcal{I}_\textrm{D}(\hat{\mathbf{s}},f) &= \frac{1}{\Nant(\Nant-1)}\,\sum_{ab}\, V_{ab}(f)\,e^{i 2\pi f\mathbf{r}_{ab}\!\cdot\,\hat{\mathbf{s}}/c},
\end{align}
where, $N_A$ is the number of antennas in the array and we have assumed equal weighting for each antenna pair.

When the array is homogeneous, $\mathcal{W}^\textrm{I} \equiv \mathcal{W}^\textrm{I}_a \equiv \mathcal{W}^\textrm{I}_b$, and the above equation reduces to:
\begin{align}\label{eqn:apndx-dirty-image-homogeneous}
  \mathcal{I}_\textrm{D}(\hat{\mathbf{s}},f) &= |\mathcal{W}^\textrm{I}(\hat{\mathbf{s}},f)|^2\,\mathcal{I}_\textrm{D}^\textrm{iso}(\hat{\mathbf{s}},f),
\end{align}
where, $\mathcal{I}_\textrm{D}^\textrm{iso}(\hat{\mathbf{s}},f)$ is the dirty image with no beam weighting (isotropic, uniform weighting of the sky) determined only by the array layout and $|\mathcal{W}^\textrm{I}|^2$ is the directional power pattern of the antenna pair familiar in standard interferometry.

In order to understand an image from a heterogeneous array, we start by looking at the contribution to the image from the visibility, $V_{ab}$, from each antenna pair. While imaging, EPIC introduces a weighting to each antenna during gridding as given by equations~\ref{eqn:dirty-image-FX} and \ref{eqn:e-field-conv}, and equivalently, the weighted visibility from an antenna pair $ab$ projected on the grid is:
\begin{align}
  V^\prime_{ab}(\mathbf{r},f) &= B^{\textrm{G}\star}_{ab}(\mathbf{r}-\mathbf{r}_{ab},f)\,V_{ab},
\end{align}
where, $B^\textrm{G}_{ab}(\mathbf{r}-\mathbf{r}_{ab},f)$ is obtained by the spatial cross-correlation of weighting kernels associated with the individual antennas using the definition:
\begin{align}
  B^\textrm{G}_{ab}(\mathbf{r}-\mathbf{r}_{ab},f) &= \int \left[ W^{\textrm{G}\star}_b(\mathbf{r}^\prime+\mathbf{r}-\mathbf{r}_b,f)\right. \nonumber\\
  &\qquad\qquad\qquad \left.\times\, W^\textrm{G}_a(\mathbf{r}^\prime-\mathbf{r}_a,f)\right]\,\dif^2\mathbf{r}^\prime.
\end{align}
The superscript `$\textrm{G}$' denotes the weighting introduced in analysis during the gridding process. Optimal imaging requires $W^\textrm{G}_a = W^\textrm{I}_a$ \citep{mor09,mor11}. However, we keep the two superscripts separate here to describe output images made assuming wrongly that the array is homogeneous.

The sky response of the weighted visibility is:
\begin{align}
  \mathcal{I}^\prime_{ab}(\hat{\mathbf{s}},f) &= \int V^\prime_{ab}(\mathbf{r},f)\,e^{i 2\pi f\mathbf{r}\cdot\,\hat{\mathbf{s}}/c}\,\dif^2\mathbf{r} \nonumber\\
                                              &= \int B^{\textrm{G}\star}_{ab}(\mathbf{r}-\mathbf{r}_{ab},f)\,V_{ab}\,e^{i 2\pi f\mathbf{r}\cdot\,\hat{\mathbf{s}}/c}\,\dif^2\mathbf{r} \nonumber\\
                                              &= V_{ab}\,\int \left[\int W^\textrm{G}_b(\mathbf{r}^\prime+\mathbf{r}-\mathbf{r}_a,f)\right. \nonumber\\
                                              &\qquad\qquad\,\,\, \left.\times\, W^{\textrm{G}\star}_a(\mathbf{r}^\prime-\mathbf{r}_a,f)\,\dif^2\mathbf{r}^\prime\right]\,e^{i 2\pi f\mathbf{r}\cdot\,\hat{\mathbf{s}}/c}\,\dif^2\mathbf{r} \nonumber\\
                                              &= \mathcal{W}^{\textrm{G}\star}_a(\hat{\mathbf{s}},f)\mathcal{W}^\textrm{G}_b(\hat{\mathbf{s}},f)\,V_{ab}\,e^{i 2\pi f\mathbf{r}_{ab}\!\cdot\,\hat{\mathbf{s}}/c},
\end{align}
which denotes the fringe from the weighted visibility. The image output of EPIC is equivalent to averaging these weighted fringes from all antenna pairs:
\begin{align}\label{eqn:apndx-wt-dirty-image-EPIC}
  \mathcal{I}^\prime(\hat{\mathbf{s}},f) &= \frac{1}{\Nant(\Nant-1)}\,\sum_{\substack{ab\\a\ne b}}\,\mathcal{I}^\prime_{ab}(\hat{\mathbf{s}},f) \nonumber\\
                                         &= \frac{1}{\Nant(\Nant-1)}\,\sum_{\substack{ab\\a\ne b}}\,\left[\mathcal{W}^{\textrm{G}\star}_a(\hat{\mathbf{s}},f)\mathcal{W}^\textrm{G}_b(\hat{\mathbf{s}},f)\right. \nonumber\\
  &\qquad\qquad\qquad\qquad\qquad\qquad \times \left. V_{ab}\,e^{i 2\pi f\mathbf{r}_{ab}\!\cdot\,\hat{\mathbf{s}}/c}\right].
\end{align}

If all antennas are identically weighted ($W^\textrm{G} \equiv W^\textrm{G}_a \equiv W^\textrm{G}_b$), using equation~\ref{eqn:apndx-dirty-image-homogeneous}, equation~\ref{eqn:apndx-wt-dirty-image-EPIC} reduces to: 
\begin{align}\label{eqn:apndx-wt-dirty-image-homogeneous}
  \mathcal{I}^\prime(\hat{\mathbf{s}},f) &= \frac{\left|\mathcal{W}^\textrm{G}(\hat{\mathbf{s}},f)\right|^2}{\Nant(\Nant-1)}\,\sum_{\substack{ab\\a\ne b}}\,V_{ab}\,e^{i 2\pi f\mathbf{r}_{ab}\!\cdot\,\hat{\mathbf{s}}/c} \nonumber\\
  &= \left|\mathcal{W}^\textrm{G}(\hat{\mathbf{s}},f)\right|^2\,\mathcal{I}_\textrm{D}(\hat{\mathbf{s}},f) \nonumber\\
  &= \left|\mathcal{W}^\textrm{G}(\hat{\mathbf{s}},f)\right|^2\,\left|\mathcal{W}^\textrm{I}(\hat{\mathbf{s}},f)\right|^2\,\mathcal{I}_\textrm{D}^\textrm{iso}(\hat{\mathbf{s}},f).
\end{align}
Thus, the dirty image, which is already attenuated by the instrumental power pattern, gets further attenuated by the power pattern introduced in the gridding step in EPIC imager. This is consistent with \citet{mor09}. 

We note that all quantities in the celestial plane (calligraphic fonts) are functions of position, $\hat{\mathbf{s}}$, and frequency, $f$. For convenience, we drop writing this dependence explicitly hereafter.

For a heterogeneous array, it is necessary to keep track of antennas and antenna pairs of different types which will determine the final weighting in the synthesized image. We consider a heterogeneous array consisting of a total of $\Nant$ antennas under $N_\textrm{T}$ different types. And under each antenna type $p$, there are $n_p$ antennas such that $\sum_p n_p = \Nant$ (unrelated to the noise quantity also denoted by $n$ in \S\ref{sec:math}). The total number of unique non-zero spacing antenna pairs is $\Nant(\Nant-1)/2$ and there are potentially up to $N_\textrm{T}(N_\textrm{T}+1)/2$ unique antenna pair types obtained by pairwise combination of the $N_\textrm{T}$ antenna types. The total number of antenna pairs, 
\begin{align}\label{eqn:apndx-total-baselines}
  \Nant(\Nant-1) &= \sum_{pq} n_{pq}, \\
  \textrm{where}, \quad   n_{pq} &= \begin{cases}
    n_p\,n_q, & p\ne q \\
    n_p(n_p-1), & p=q, 
  \end{cases}
\end{align}
is obtained by counting antenna pairs, $n_{pq}$, in different antenna pair types, $pq$. Note that $ab$ and $pq$ act as simple indices into pairs formed from individual indices and not their product.

Since equation~\ref{eqn:apndx-wt-dirty-image-EPIC} is obtained by averaging weighted fringes over all antenna pairs, it can be expressed as: 
\begin{align}\label{eqn:apndx-wt-dirty-image-EPIC-decomp}
  \mathcal{I}^\prime &= \frac{1}{\Nant(\Nant-1)}\,\sum_{pq}\sum_{\substack{ab\\a\ne b\\a\in p, b\in q}}\,\mathcal{I}^\prime_{ab}
\end{align}
Here, equation~\ref{eqn:apndx-wt-dirty-image-EPIC} has been re-written as the sum of fringes over all antenna pairs in an antenna pair type (inner sum) and subsequently summed over all antenna pair types (outer sum). The effective attenuation of the dirty image due to instrumental and gridding weights is:
\begin{align}\label{eqn:apndx-effective-weighting}
  \mathcal{W}_\textrm{eff} &= \mathcal{I}^\prime\, / \,\mathcal{I}_\textrm{D}^\textrm{iso}.
\end{align}

We can simplify further if we assume that the spatial distribution of antenna pairs in each antenna pair type are similar and thus result in a similar dirty image, $\mathcal{I}_\textrm{D}^\textrm{iso}$. This is valid when antennas under each antenna type are chosen such that differences in the point spread functions are insignificant. Then, equation~\ref{eqn:apndx-wt-dirty-image-EPIC-decomp} can be written as:
\begin{align}\label{eqn:apndx-wt-dirty-image-EPIC-decomp-approx1}
  \mathcal{I}^\prime &\approx \frac{\mathcal{I}_\textrm{D}^\textrm{iso}}{\Nant(\Nant-1)}\,\sum_{pq}\sum_{\substack{ab\\a\ne b\\a\in p, b\in q}} \mathcal{W}^{\textrm{G}\star}_a\,\mathcal{W}^\textrm{G}_b\,\mathcal{W}^\textrm{I}_a\,\mathcal{W}^{\textrm{I}\star}_b,
\end{align}
where, $\mathcal{I}_\textrm{D}^\textrm{iso}$ is the same for all terms and has been pulled outside the summations. For all antennas indexed by $a$ that belong to a particular type $p$, we replace $\mathcal{W}_a$ with $\mathcal{W}_{(p)}$ and is applicable to those arising from both instrumental (superscript `$I$') and gridding (superscript `$G$') origins. Using arguments similar to those in equation~\ref{eqn:apndx-total-baselines}, equation~\ref{eqn:apndx-wt-dirty-image-EPIC-decomp-approx1} can be further simplified to:
\begin{align}\label{eqn:apndx-wt-dirty-image-EPIC-decomp-approx2}
  \mathcal{I}^\prime &\approx \frac{\mathcal{I}_\textrm{D}^\textrm{iso}}{\Nant(\Nant-1)}\,\sum_{pq}n_{pq}\,\mathcal{W}^{\textrm{G}\star}_{(p)}\,\mathcal{W}^\textrm{G}_{(q)}\,\mathcal{W}^\textrm{I}_{(p)}\,\mathcal{W}^{\textrm{I}\star}_{(q)}.
\end{align}

This means $\mathcal{W}_\textrm{eff}$ can be expressed as fourth-power combinations of antenna voltage patterns each weighted by the number of antenna pairs in those antenna pair types:
\begin{align}
  \mathcal{W}_\textrm{eff} &\approx \frac{1}{\Nant(\Nant-1)}\,\sum_{pq}n_{pq}\,\mathcal{W}^{\textrm{G}\star}_{(p)}\,\mathcal{W}^\textrm{G}_{(q)}\,\mathcal{W}^\textrm{I}_{(p)}\,\mathcal{W}^{\textrm{I}\star}_{(q)}.
\end{align}
In an optimally weighted image, $\mathcal{W}^\textrm{G}_{(p)} \equiv \mathcal{W}^\textrm{I}_{(q)}$. Then, equation~\ref{eqn:apndx-effective-weighting} reduces to:
\begin{align}\label{eqn:apndx-effective-weighting-optimal}
  \mathcal{W}_\textrm{eff}^\textrm{opt} &\approx \frac{1}{\Nant(\Nant-1)}\,\sum_{pq} n_{pq}\,\left|\mathcal{W}^\textrm{I}_{(p)}\right|^2\left|\mathcal{W}^\textrm{I}_{(q)}\right|^2,
\end{align}
where the superscript `$\textrm{opt}$' denotes `optimal'.




\bsp	
\label{lastpage}
\end{document}